# A community-powered search of machine learning strategy space to find NMR property prediction models

Lars A. Bratholm[1,2]*, Will Gerrard[1], Brandon Anderson[21], Shaojie Bai[5,6], Sunghwan Choi[7], Lam Dang[8], Pavel Hanchar[9], Addison Howard[10], Guillaume Huard[21], Sanghoon Kim[11], Zico Kolter[5,6], Risi Kondor[12,13,14], Mordechai Kornbluth[15], Youhan Lee[16], Youngsoo Lee[17], Jonathan P. Mailoa[15], Thanh Tu Nguyen[8], Milos Popovic[18], Goran Rakocevic[18], Walter Reade[10], Wonho Song[19], Luka Stojanovic[18], Erik H. Thiede[12,14], Nebojsa Tijanic[18], Andres Torrubia[20], Devin Willmott[5], Craig P. Butts,[1] David R. Glowacki[1,3,4]*
& Kaggle participants[21]

[1]School of Chemistry, University of Bristol, Cantock's Close, Bristol BS8 1TS, UK; [2]School of Mathematics, University of Bristol, Fry Building, Woodland Road, Bristol BS1 1UG, UK; [3]Dept. of Computer Science, University of Bristol, Merchant Venturer's Building, Bristol BS8 1UB, UK; [4]Intangible Realities Laboratory, University of Bristol, Cantock's Close, Bristol BS8 1TS, UK; [5]Bosch Center for Artificial Intelligence, Pittsburgh, PA 15222, USA; [6]Carnegie Mellon University, Pittsburgh, PA 15213, USA; [7]National Institute of Supercomputing and Network, Korea Institute of Science and Technology Information, 245 Daehak-ro, Yuseong-gu, Daejeon 34141, Republic of Korea; [8]BNP Paribas Cardif; [9]Fyusion, Inc.; [10]Kaggle, Google Inc., Mountain View, US; [11]Ebay Korea, 34F, Gangnam Finance Center, 152 Teheran Ro, Gangnam Gu, Seoul, Republic of Korea; [12]Department of Computer Science, The University of Chicago, 5730 S Ellis Ave, Chicago, IL 60637; [13]Department of Statistics, The University of Chicago, 5747 S Ellis Ave, Chicago, IL 60637; [14]Center for Computational Mathematics, Flatiron Institute, 162 5th Ave., New York, NY 10010; [15]Bosch Research and Technology Center, Cambridge, MA 02139, USA; [16]Department of Chemical and Biomolecular Engineering, Korea Advanced Institute of Science and Technology, 291 Daehak-ro, Yuseong-gu, Daejeon 34141, Republic of Korea, current address: Korea Atomic Energy Research Institute, (34057) 111, Daedeok-daero 989beon-gil, Yuseong-gu, Daejeon, Republic of Korea; [17]MINDS AND COMPANY, 2621, Nambusunhwan-ro, Gangnam-gu, Seoul 06267, Republic of Korea; [18]Totient Inc, Sindjeliceva 9, 11000 Belgrade, Serbia; [19]KAIST Web Security & Privacy Lab, 291, Daehak-ro, Yuseong-gu, Daejeon 34141, Republic of Korea; [20]Medbravo.org, Alicante, Spain; [21]Participants are listed at www.kaggle.com/c/champs-scalar-coupling/leaderboard;

* lars.bratholm@bristol.ac.uk, drglowacki@gmail.com

*The rise of machine learning (ML) has created an explosion in the potential strategies for using data to make scientific predictions. For physical scientists wishing to apply ML strategies to a particular domain, it can be difficult to assess in advance what strategy to adopt within a vast space of possibilities. Here we outline the results of an online community-powered effort to swarm search the space of ML strategies and develop algorithms for predicting atomic-pairwise nuclear magnetic resonance (NMR) properties in molecules. Using an open-source dataset, we worked with Kaggle to design and host a 3-month competition which received 47,800 ML model predictions from 2,700 teams in 84 countries. Within 3 weeks, the Kaggle community produced models with comparable accuracy to our best previously published 'in-house' efforts. A meta-ensemble model constructed as a linear combination of the top predictions has a prediction accuracy which exceeds that of any individual model, 7-19x better than our previous state-of-the-art. The results highlight the potential of transformer architectures for predicting quantum mechanical (QM) molecular properties.*

## 1. Introduction

The rise of machine learning (ML) in the physical sciences has created a number of notable successes, (*1-7*) and the number of published outputs is increasing substantially. (*8*) This explosion is perhaps not entirely surprising, given that ML 'search space' is effectively infinite. For example, the performance of a particular ML algorithm strategy depends sensitively on at least four components: (a) the dataset used for training (and the corresponding methodology used for dataset curation); (b) the feature selection used to construct ML inputs; (c) the choice of ML algorithm; and (d) the values of the optimal constituent hyperparameters. For components (b) and (c), the space of possibilities is continually expanding; for components (a) and (d), the space of possibilities is potentially infinite. Given the sensitivity of ML approaches to each of the items outlined above, ML's explosion within the scientific literature has led to warnings of an emerging computational reproducibility crisis, a risk exacerbated by the fact that many peer-reviewed ML publications do not include the data and algorithms required to reproduce their results. (*9*)

The difficulty of searching an enormous ML space is compounded by the fact that the training of even simple neural networks has been shown to be an NP-complete problem. (*10*) Deciphering whether any global optima lurk within an effectively infinite ML search space has been the topic of a great deal of research; however, there seems to be a consensus emerging that it is practically impossible to demonstrate that any particular ML strategy is in fact optimal or bias-free, even for very simple systems. (*11*) Broadly speaking, the parameter spaces in which a particular ML strategy can be constructed are non-convex, and characterized by multiple local minima and saddle points in which optimization algorithms can get trapped. (*12*) Nevertheless, ML algorithms can produce useful results. In a nod to the 1950 Japanese period drama "Rashomon" (where various characters provide subjective, alternative, self-serving, yet compelling versions of the same incident), ML's tendency to produce many accurate-but-different models has been referred to as the "Rashomon effect" in machine learning. (*13*) In such a vast space, any individual agent has a chance of stumbling upon a reasonable ML model. Given the difficulty of rationalizing the uniqueness of any particular ML model or approach, individual models are increasingly being used as constituents within ensemble models, whose combined accuracy outperforms that of any individual model. (*14*)

Over the last several years, a number of studies have demonstrated the utility of 'crowd-sourced' approaches for solving scientific problems which involve searching hyperdimensional spaces. (*15-19*) Inspired by recent attempts within both particle physics (*20, 21*) and materials science (*22*) using community power to develop ML algorithms, we worked with Kaggle (an online platform for ML competitions), to design a competition encouraging participants to develop ML models able to accurately predict QM nuclear magnetic resonance (NMR) properties from 3D molecular structure information. (*23*) The fact that some of our authorship team had worked in this area over several years (*24*) meant that we had quantitative and qualitative benchmarks to analyse competition progress in relation to what conventional academic research approaches had achieved. The so-called 'Champs Kaggle Competition' (CKC) ran from 29-May-19 through 28-Aug-19. The 5 models which achieved the highest accuracy were awarded respective prize money of $12.5k, $7.5k, $5k, $3k and $2k. Over ~13 weeks, the CKC received 47,800 model predictions from 2,700 teams in 84 countries (Fig 1a), representing the most exhaustive search to date of ML strategies aimed at predicting QM NMR properties from 3D molecular structure information. The number of participants who engaged with the CKC was amongst the highest for any physical science challenge which Kaggle has hosted to date. Fig 1b and 1c show a steady increase in the number of participants who joined the CKC versus time. CKC participants reported being drawn to the competition because it: (a) facilitated progress on an important research problem; (b) involved a rich, noise-less dataset whose structure was easy to understand; and (c) had a dataset which was manageable using standard data processing tools, workflows, and hardware.

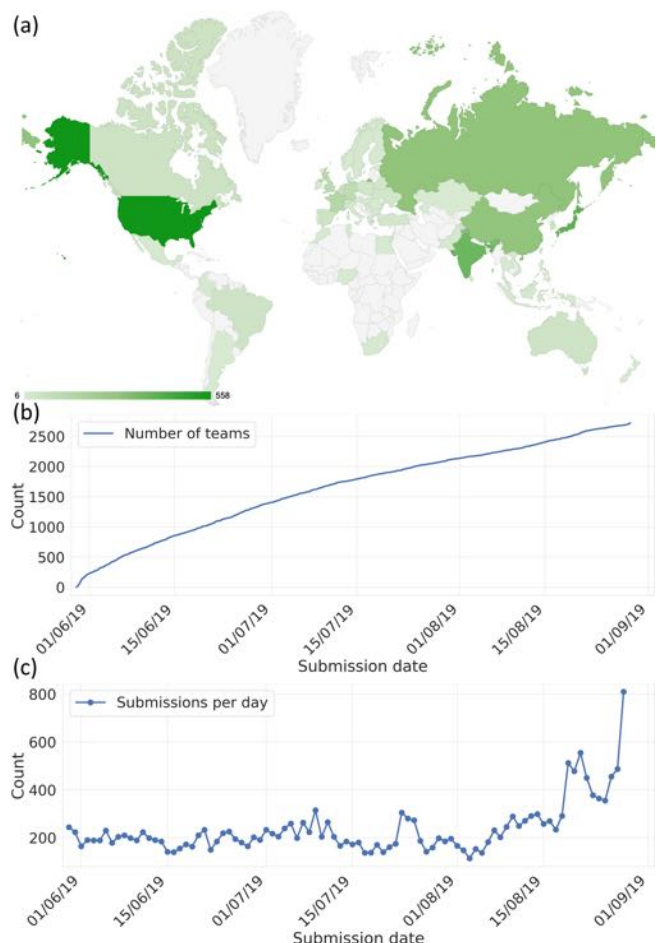

**Figure 1**: (a) map showing the number teams participating from different countries over the duration of the CKC (countries with less than 5 participants are shown light gray); (b) the number of CKC participants vs. time; (c) number of submissions per day

## 2. Competition Design
### 2.1. Domain

NMR is the dominant spectroscopic technique for determining 2D and 3D molecular structure in solution. Amongst the most important data obtained in an NMR spectrum are the chemical shifts (which describe the position/frequency of a signal in the spectrum) and the scalar couplings (which determine the splitting/shape of the signal in the spectrum). ML methods to predict NMR properties are established in academic and commercial workflows for determining 2D molecular structure from experimental NMR datasets. (*25-28*) Despite this success, these 2D approaches often fail when the NMR properties are affected by 3D structure, for example atoms are separated by several bonds yet remain close in 3D space (ring current effects, hydrogen bonding etc.). This is an inherently difficult problem as the 3D molecular structure is simply not well described by 2D representations and there are not enough high-quality experimental data available to accurately infer most 3D relationships from a 2D structural representation alone.

The most accurate computed predictions of NMR properties use QM methods like density functional theory (DFT) to get a one-to-one mapping between a 3D structure and the contribution it has to the experimentally observed NMR property. Accurate QM methods for NMR property predictions are powerful but expensive. Recent work has thus focussed on developing ML algorithms which can efficiently reproduce the results of costly QM methods, achieving results in seconds rather than hours or days. (*24, 29*) ML approaches have the added appeal that they can be trained using large datasets of DFT-computed NMR parameters, which are not limited to experimental structural observations. With a large enough training database, we have shown in previous work that an ML strategy can approach the accuracy of DFT calculations of atom-centered NMR parameters such as chemical shift for 3D structure analysis, but with several orders of magnitude reduction in time. (*24*)

Beyond NMR, the last decade has seen considerable effort focused on machine learning QM molecular properties. (*30-36*) Broadly speaking, this work has tended to focus on predicting atomic properties such as partial charges, or molecular properties such as energies and dipoles. Relatively little work has been carried out designing ML models which are able to predict pairwise atomic properties such as scalar coupling constants. Our earlier work to develop pairwise property prediction algorithms were effectively independent-atom treatments, in which atomic feature vectors describing the local environment of each atom were concatenated. (*24*) However, this approach loses information about the relative position/orientation of each atom's respective environment, which is important for multiple-bond couplings. The CKC represents an attempt to kickstart research into ML methods able to make accurate prediction of pairwise properties.

### 2.2. Dataset & Scoring

Scalar couplings are critically dependent on the 3D structure of the molecule for which they are being measured; however at the time we carried out this work, we were unaware of accurate experimental databases linking pair-wise mutiple-bond NMR scalar couplings to well-defined 3D molecular structures. Therefore, we decided to run the CKC utilizing molecular structures included in the QM9 dataset, a publicly available benchmark for developing ML models of 3D structure-property relationships. (*37*) QM9 includes ~134k molecules comprised of carbon, fluorine, nitrogen, oxygen and hydrogen. The molecules included within QM9 have no more than 9 heavy atoms (non-hydrogen), with a maximum of 29 total atoms. To obtain a corresponding set of scalar couplings, we extended the QM9 computational methodology, using the B3LYP functional (*38*) and the 6-31g(2df,p) basis set (*39-42*) to compute NMR parameters on the optimized QM9 structures. The computed QM9 scalar coupling constants are available under Creative Commons CC-NC-BY 4.0, enabling others to build on this work.

To remove the possibility of CKC participants overfitting their models to the entire set of computed QM9 scalar couplings, 65% of molecules in the dataset were randomly partitioned into a training set and the other 35% to a testing set. The test set was further split, with 29% of the data in a 'public' test set, and 71% of the data in a 'private' test set (competitors were unaware of the specifics of the private/public split). Both the training and test sets included the molecular geometries and indices of the coupling atoms. Unlike the test set, the training set included a range of other data, including the calculated scalar coupling values, their breakdown into Fermi contact (FC), spin-dipole (SD), paramagnetic spin-orbit (PSO) and diamagnetic spin-orbit (DSO) components, and a range of auxiliary information obtained from the QM computations (e.g., potential energy, dipole moment vectors, magnetic shielding tensors and Mulliken charges). As the CKC progressed, participating teams continually iterated and improved their models. A regularly updated and publicly visible leaderboard enabled each team to see where their model ranked in predicting the public test set data compared to the model predictions made by all of the other teams.

The leaderboard scores were determined using a function which accounted for the 8 different types of coupling constants included in the training and testing datasets: $^1J_{HC}$, $^1J_{HN}$, $^2J_{HH}$, $^2J_{HC}$, $^2J_{HN}$, $^3J_{HH}$, $^3J_{HC}$ and $^3J_{HN}$ (where the superscript indicates the number of covalent bonds separating the atom pairs indicated by the subscript). Since the number of couplings of each type differed (e.g., the molecular composition of the QM9 test set included 811,999 $^3J_{HC}$ couplings compared to 24,195 $^1J_{HN}$ couplings) and spanned different value ranges, the scoring function used the average of the logarithm of the mean absolute error for each type of coupling constant:

$$\text{score} = \frac{1}{T}\sum_{t}^{T} \log\left(\frac{1}{n_t}\sum_{i}^{n_t} |y_i - \hat{y}_i|\right) \qquad \text{Eq (1)}$$

where $t$ is an index that runs over the $T = 8$ different scalar coupling types, $i$ is an index that spans $1..n_t$, the number of observations of type $t$, $y_i$ is the scalar coupling constant for observation $i$, and $\hat{y}_i$ is the predicted scalar coupling constant for observation $i$. This scoring function ensures, for example, that a 10% improvement in one type of coupling will improve the score by the same amount as a 10% improvement in another type of coupling, so that no coupling class dominates.

## 3. Results
### 3.1 Leaderboard time evolution

Over the course of the CKC, Fig 2a shows the evolution of the best score whose source code was publicly available (public notebooks), and its relationship to the top score versus time. Fig 2b shows that the time trace of the top score is well fit by a bi-exponential curve with two distinct phases. **Phase 1** lasted for the first week, during which time the accuracy increased by ~12x (~2.5 improvement in score), with a time constant of ~1.29 ± 0.18 days. **Phase 2** lasted for the next 12 weeks, during which time the accuracy improved more gradually by a factor of ~4x (~1.5 improvement in score), with a time constant of ~50.0 ± 16.6 days. To determine which models were awarded prize money, the final set of model rankings were assessed using Eq (1) to evaluate how well each of the models predicted the scalar coupling values in the private test set (preventing competitors inferring the target property from the leaderboard scores rather than from the training set). Due to the large amount of noise-less data, the positioning in the top 37 submissions was the same on the public and private leaderboard at the end of the CKC. Several teams commented that the stability between the public & private leaderboards made for an enjoyable competition.

The top-scoring method achieved a geometric mean error (exponential of the score) of 0.039 Hz which was 6-16x more accurate than what could be achieved using our own recently developed methodology (see SI for details). (*24*) In addition to the final score, Kaggle also rewards participants who make the best contributions to: (1) publicly available code, and (2) the discussion forums. As a result of these incentives, a number of participants opted to voluntarily publish their source code (public notebooks). In many cases, the public notebooks were then utilized and adapted by other CKC participants. As shown in Fig 2a, the best score achieved using these public notebooks follows a time trace which is similar to the leading score, but less accurate by ~1.5. A number of participants made instructional web posts, scripts, and videos outlining specific approaches which they had taken during the CKC. For example, video presentations by Andrey Lukyanenko (*43*) and the NVIDIA team (*44*) discuss the approaches which they utilized to develop the 8[th] and 33[rd] place solutions, respectively. The CKC summary features insightful write-ups by several top teams in which they describe their various model approaches. (*45*)

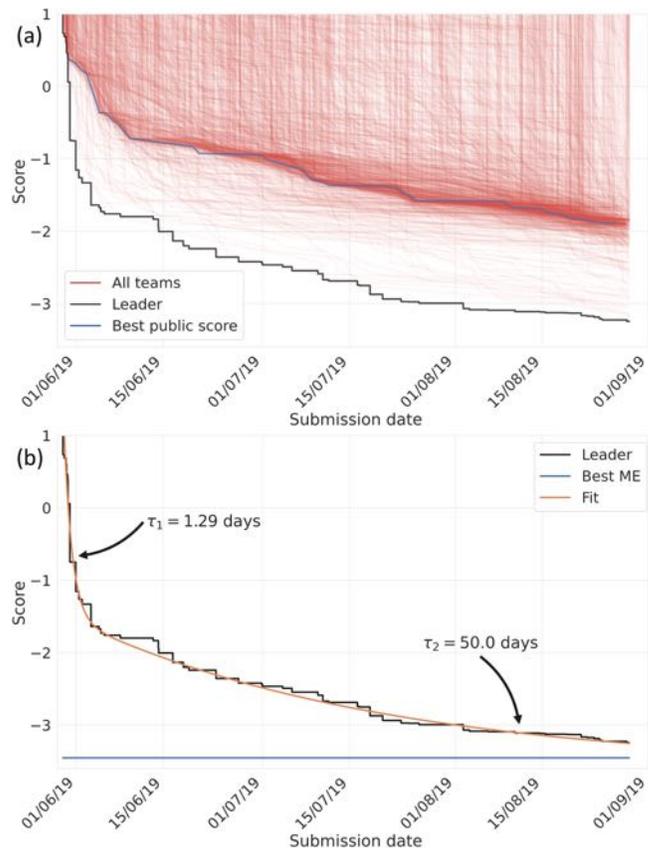

Figure 2: (a) score evolution vs. time. Black line shows the best performing method vs. time. Blue line shows the best performing public notebook. Red lines shows the best submission by each team; (b) best fit of the time dependent leader (black) score to an biexponential curve of the form $A \cdot \exp(-t/\tau_1) + B \cdot \exp(-t/\tau_2) + C$ ($A = 2.11$; $B = 2.97$; $\tau_1 = 50.0$ days; $\tau_2 = 1.29$ days; $C = -3.59$). Blue indicates the best ME model score.

### 3.2 Meta-Ensemble Model

To assess the extent to which the prize-winning submissions differed from one another (and other highly ranked submissions), we used the top 400 submissions to construct a meta ensemble (ME) model as a linear combination of the top scoring models:

$$y_{i,ME} = \sum_{j=k}^{400} w_j y_{i,j} \qquad \text{Eq (2)}$$

Given that many of the top models (and all of the prize winners) were ensemble models, we have adopted the term "meta-ensemble" (ME) to emphasize the fact that Eq (2) is an ensemble of ensemble models. In Eq (2), the ME prediction $y_{i,ME}$ of the $i$'th scalar coupling constant is a linear combination of the predictions $y_{i,j}$ of the $j$'th ranked model. The index $k$ specifies the lowest ranked model to be included within the optimized ME model. When $k = 1$, Eq (2) runs over the entire list of the top 400 models. When $k > 1$, Eq (2) neglects top-scoring models. Setting $k = 6$ for example, the Eq (2) ME model excludes all of the prize-winning models (ranks #1 – #5). For ME models constructed using Eq (2), the weights $w_j$ were determined by minimizing $y_{i,ME}$ using half of the test set, under the constraint that the weights were positive and summed to unity. While a range of different ME models can be constructed (e.g., different ensembles for each type of coupling, median averaging etc.), this simple mean is easy to interpret.

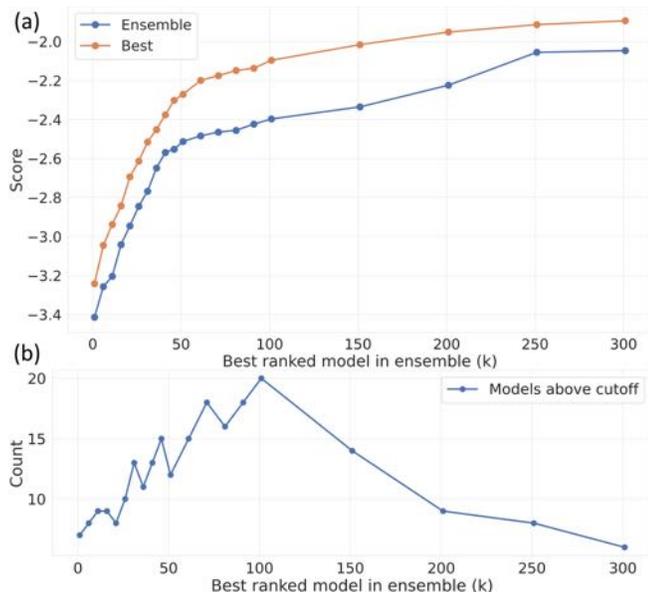

Figure 3: (a) comparison of the top individual score (orange) to the ME model score (blue) as function of the k value in Eq (2); (b) the number of contributors to the ME model at a particular k value that had an optimized weight greater than 0.01.

Different classes of machine learning algorithms (or even the same algorithm with different hyperparameters) may be able to learn different regions of the data better than others. Thus, by combining the highest scoring model predictions that have the least correlation for a meta-ensemble, the strengths of various models may be accumulated, a result confirmed by the ME analysis shown in Fig 3a as a function of $k$. As expected (for $k = 1..300$) the optimized ME model achieves an accuracy which always surpasses that of the best individual model. In the regime where the top scorers are incrementally being eliminated from Eq (2) ($k = 1..50$), Fig 3a shows that the ME model has a score that is ~0.2 lower than the "best" model. For example, the $k = 7$ ME model (which neglects the top 6 models) still outperforms the winning solution, and the $k = 11$ ME model outperforms the winning solution when the per-type ensemble mentioned above is used. Fig 3b shows how many contributors to the ME model at a particular k value had an optimized weight greater than 0.01. Broadly speaking, the Fig 3a results can be lumped into three regimes. In the first regime ($k \sim 1..40$) the best performing methods dominate the ME and there is little to be gained by including within the ME methods that are very different if they perform worse. In the second regime ($k \sim 41..200$), Fig 3a shows that the gap between the top score and the ME model widens to ~0.4. Here there are many similarly performing yet different methods, so there is much to be gained by combining their different approaches into a ME. In the third regime ($k \sim 201..300$) the gain from a ME decreases, presumably because many of the models are similar variants of the public notebooks. The relative benefit of constructing a ME model (versus using a top-scoring model) thus appears to be more significant outside of the band of top-scoring and low-scoring models.

For the $k = 1$ ME model, which was 7-19x more accurate than our previously published model, (24) we analysed in further detail its constituents. The results in Table 1 show the $k = 1$ ME constituents with weights $w_i > 0.02$, along with the relative rankings $j$ of the constituent ME models. Table 1 shows that there is no particular model which is dominant: there are five models with a weighting greater than 0.11, and three with a weighting greater than 0.20. Of the six models in Table 1, one (#12) falls outside the top 5. Its 0.149 contribution is larger than prize winning models #3 and #5. Fig 4a shows the submission history of the Table 1 models, and their relationship to the overall public leader board.

Table 1: Summary descriptions for the six models in the final ME. "Use of Scalar coupling components" refers to whether a team decomposed the scalar couplings into four separate components in their model.

|  | 1st | 2nd | 3rd | 4th | 5th | 12th |
|---|---|---|---|---|---|---|
| Weight $w_j$ | 0.204 | 0.270 | 0.111 | 0.203 | 0.046 | 0.149 |
| Number of submissions | 73 | 167 | 151 | 37 | 53 | 4 |
| Country | USA | Spain Belarus | S. Korea | Serbia | France | USA |
| Team size | 5 | 2 | 5 | 4 | 3 | 2 |
| Any Chemistry expertise? | Y | N | Y | Y | N | Y |
| Use of scalar coupling components? | N | N | Y | N | N | N |
| Translational invariance? | Y | Y | N | Y | Y | Y |
| Rotational invariance? | Y | N | N | Y | Y | Y |
| Previous Kaggle experience? | N | Y | Y | N | Y | N |
| Included additional input features? | Y | N | N | Y | Y | N |
| Number of model parameters | ~105M | ~60M | ~70M | ~60M | ~66M | ~250K |

### 3.3 Correlation Analysis

To further understand the relationship between the winning submissions within the $k = 1$ ME model, we carried out a correlation analysis on the top 50 team submissions. The submissions were then ordered using a hierarchical clustering analysis (see SI). The results in Fig 4b show that the #1 – #5 teams are part of the same sub-cluster i.e. all relatively similar to each other. Fig 4c specifically highlights the low correlation between models #1 – #5 compared to model #12, which shows that this team's approach exists within a region of ML strategy space that appears relatively distinct from the prize-winning models, and also from the top 50 solutions.

Compared to the others in Table 1, team #12 was a relative latecomer to the CKC as shown in Fig 4a. In addition, the number of parameters in their model is ~100x smaller than the others. The low correlation of team #12 compared to the other teams in Table 1 appears to have arisen because they utilized the 'Cormorant' rotationally covariant neural network strategy. (46) Originally developed for learning molecular potential energy surfaces (PESs), Cormorant takes advantage of rotational symmetries in order to enforce physical relationships in the resultant neural network, by using spherical tensors to encode local geometric information around each atom's environment, which transform in a predictable way under rotation. The use of spherical tensors allows for a network architecture that is covariant to rotations, so that if a rotation is applied to a layer, all activations at the next layer will automatically inherit that rotation. As such, a rotation to a Cormorant input will propagate through the network to ensure that the output transforms as well. This captures local geometric information while still maintaining the desired transformation properties under rotations. Team #12's sophisticated input processing strategy contrasts with the approaches taken by other teams, which tended to utilize far simpler encoding strategies, either by restricting the input features to have translational and rotational invariance (e.g., using internal distances), adding translational and rotational noise to make the inputs robust to rotation and translation, or allowing the model learn invariance on its own. Team #12's approach is grounded in domain specific physics knowledge, and characteristic of the emphases which physical scientists tend to apply in ML contexts.

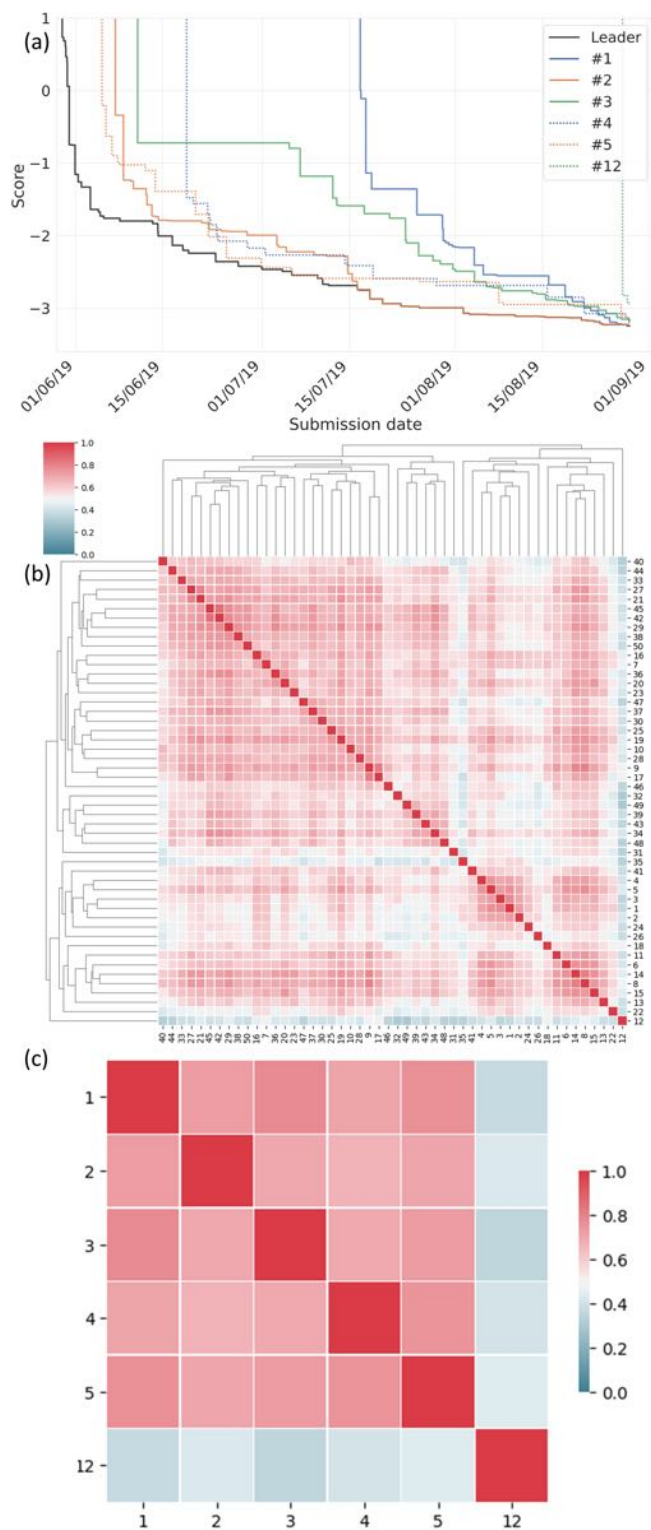

Figure 4: (a) score evolution vs. time for Table 1 teams. Black line shows the best performing method at a current time. Colored lines show the best submission by each team; (b) correlation amongst the top 50 submissions. Red indicates high correlation, and blue low. Bottom and right side shows the ranking of the submission, while top and left features a dendrogram depicting the hierarchical clustering; (c) correlation between Table 1 teams.

## 4. Discussion & Conclusions

Community-powered approaches offer a powerful tool for searching ML strategy space and providing accurate predictions for physical science problems like the prediction of 2-body QM NMR properties. Within 3 weeks, the best score on the Kaggle public leader board achieved an accuracy which surpassed our own previously published approaches, (*24*) suggesting that an open source community-powered 'swarm search' of ML strategy space may in some cases be significantly faster and more cost-efficient than conventional academic research strategies where a single agent (e.g., a PhD student or post-doctoral researcher) spends several years hunting for solutions in an infinite search space. ME model construction combined with correlation analysis highlights the strength of the CKC 'swarm search' approach, in line with the "Rashomon effect".

Whereas our earlier approaches to predicting NMR structure coupling constants (*24*) had relied on kernel-ridge regression approaches (*47*) where the internal distances and angles in the molecules were systematically encoded to a feature vector for the coupling atom pairs using predetermined basis functions, the community which emerged around the CKC pioneered a new application of transformer neural nets (*48*) to QM molecular property prediction. While such networks have found extensive use for sequence modelling and transduction problems such as language modelling and machine translation, they represent a relatively new approach to predicting QM properties like NMR shifts or scalar couplings, and it will be interesting to explore their further application to other QM properties and more general 2-body property prediction problems, which are relevant in several domains across the physical sciences. The rich portfolio of open source blog posts, data, insight, source code, and discussions arising from the CKC offers an excellent foundation for subsequent research and follow-up studies, through community initiatives or more conventional academic research approaches.

Teams #2 and #5 had no domain specific expertise, and yet outperformed participants with domain expertise, including our own previous attempts. (*24*) This contrasts with previously published Kaggle competitions in particle physics (*20, 21*) and materials science, (*22*) where the winners tended to be domain experts. Table 1 shows that teams with prior domain expertise (e.g., #1 and #4) used their insight to calculate additional input features beyond those which we provided, and which they then used as model input. For example, team #1 used Mulliken charges and atomic valency, while team #4 used electronegativity, first ionization energy, electron affinity, mulliken charge, and bond types. Despite this added complexity, team #1 only narrowly managed (i.e., within the CKC's final hours) to improve on the approach of team #2, which used a simple cartesian input representation with no additional data.

All of the prize winning teams utilized deep neural networks where the encoder learned the pair-feature vectors from the coordinates, atom types, distances, etc. A separate feed forward neural network (decoder) was then used to make scalar coupling predictions per coupling type or sub coupling type. The relatively simple input descriptions used by many of the top teams transferred to the neural network the challenge of learning an effective input representation. Such approaches contrast with those favored by physical scientists, which utilize more complex descriptors constructed so as to include domain specific insight (e.g., rotational symmetries for team #12). Taking advantage of the variance in approaches, the various model predictions can be combined into a ME model whose combined accuracy surpasses that of any individual model, 7-19x more accurate than what our previous methods were able to achieve. The benefit of a ME model seems to be most significant in the regime where there are many independent individual models with similar performance.

Fig 2a shows that the average benefit which new models contributed to the overall improvement in prediction accuracy decreased versus time, with a rapid improvement over the first week, followed by a much more gradual improvement over the next 13 weeks. Fig 1c shows that the number of model predictions was approximately constant versus time with an increase over the final 20 days. These observations indicate an overall decrease in the relative cost/benefit ratio as a function of time. This cost/benefit decrease is qualitatively compatible with conclusions drawn from previous meta-analyses of scientific progress, (49) which suggest that search strategies for scientific discovery tend to become less efficient with time. In our case, these results suggest that a shorter competition may have furnished similar insights. The results also highlight potential shortcomings in the elaborate scheme of awards and prizes which scientific disciplines utilize to incentivize progress and recognize 'top-performers' – e.g., the fact that solution #12 played a more important role in the optimized ME model compared to some of the prize winning models offers an important reminder that scientific progress is a community effort that depends on a range of important contributions, which can often go unrecognized in conventional indicators of prestige.

The results of this study demonstrate how community science initiatives in conjunction with open data can enable rapid scientific progress in ML domains, reaffirming the community benefits that can arise when scientific workers make their data and algorithms open. Web-based platforms enable distributed community efforts to build engagement with scientific concepts at a time where scientific approaches face mounting challenges across media and political landscapes. Given the constraints on conventional scientific collaboration which have arisen as a result of social distancing, distributed scientific community efforts like these may become more prevalent in the near term. For example, there has been a steady increase in the number of scientific stack exchanges, which (like Kaggle) incentivize scientific communities to share knowledge and expertise. Digital platforms which benefit from the ubiquity of cloud computing and which enable distributed communities to engage with one another to undertake collective problem solving are likely to play an important role in our emerging scientific future. Such approaches may be particularly useful for problems like ML, where the strategy spaces are effectively infinite. Moving forward, it will be interesting to explore the extent to which search efficiencies might be enhanced by combining the intelligence of human agents with machine agents.

**Contributions**
LAB and DRG devised the project concept, conceived various analysis strategies, and organized overall execution of the project work strands (LAB early phases, DRG late phases). LAB, WG, DRG, CPB, AH and WR worked together to design the form of the CKC. LAB computed and curated the CKC data, managed the CKC, responded on the forums, engaged with the winners, and organized a discussion workshop. LAB carried out data analyses with assistance of WR and WG, and guidance from DRG. AH and WR mounted the project on the Kaggle platform, and advised on how to run the CKC. The following CKC participants contributed to discussions which fed into the paper, as well as contributed to solution code and write-ups which are included in the SI: ZK, DW, JPM, MK & SB (team #1); AT & PH (team #2); SC, SK, Youhan Lee, Youngsoo Lee & WS (team #3); GR, MP, NT & LS (team #4); LD, GH & TTN (team #5); and BA & ET (team #12). DRG wrote the initial paper draft based on the data analysis, with subsequent input from LAB, CPB, and WG. DRG and CPB organized project funding.

**Competing Interests:** This competition was made possible through financial support from Kaggle, where AH and WR are employees.

**Data and materials availability:** All data needed to evaluate the conclusions in the paper are present in the paper and/or the Supplementary Materials. Additional data related to this paper may be requested from the authors.

**Acknowledgments.**
Part of this work was carried out using the computational facilities of the Advanced Computing Research Centre, University of Bristol. WG is partially supported by the EPSRC National Productivity Investment Fund (NPIF) for Doctoral Studentship funding. LAB thanks the Alan Turing Institute under the EPSRC grant EP/N510129/1. DRG is supported by the Leverhulme Trust (Philip Leverhulme Prize) and Royal Society (URF/R/180033). LAB and DRG acknowledge support of this work through the "CHAMPS" EPSRC programme grant EP/P021123/1. SC was supported by National Research Foundation of Korea (2018R1D1A1B07049981, 2019M3E5D4065968) funded by the Ministry of Science and ICT. We furthermore thank Prof. Jan H. Jensen for access to additional computing resources and Dr. Christopher Sutton for helpful advice on the design of the CKC.

# Supporting Information for
# A community-powered search of machine learning strategy space to find NMR property prediction models

## S1 Overview

Section S2 and S3 contains details on how the dataset was generated, and what considerations went into the choices made. Section S4 describes the data set available to the partitipants, as well as details on how submissions were scored. Section S5 contains various plots and analysis strategies that didn't make it into the main paper, as well as details on how the ensembles were fitted and the correlations were calculated.

Section S7, S8, S9, S10, S11 and S12 contain detailed write-ups explaining the model architecture of the five winning teams as well as the 12th placed team. Finally, section S13 outlines the code and data available for download, including all the winning solutions.

## S2 Dataset considerations

We opted to use structures from the QM9 dataset[1] to construct our own dataset as well as the same computational methodology. QM9 consists of around 130,000 molecules comprised of carbon, fluorine, nitrogen, oxygen and hydrogen. Each molecule has up to 9 heavy atoms (non-hydrogen) and up to 29 total number of atoms. Each structure was optimized by the authors using Density functional theory (DFT), with the B3LYP functional[2] and the 6-31g(2df,p) basis set[3, 4, 5, 6]. There were several advantages of basing our dataset on QM9, as well as several possible disadvantages that needed to be considered. QM9 has historically had an important role as a benchmark dataset in the development of machine learning models for 3D structure-property relationships, and we deemed it beneficial to augment this with magnetic properties. This is particularly true for pairwise properties like scalar coupling constants, as there to our knowledge currently does not exist any dataset that includes pair-wise properties. Using an existing set of structures also eased the workload of generating a dataset considerably. We initially had two primary concerns about using

QM9 structures specifically and one primary concern about using computational methods in general.

- The QM9 molecules are quite small compared to molecules commonly used as e.g. medicines, and the solutions to the competition might not work (due to memory use etc.) on larger molecules. However, since the strength of atomic interactions decay with distance (with the exception of large aromatic systems), we believe that we could ultimately modify the solutions to scale better with system size by introducing an interaction distance cutoff. At this time, we have not researched the current size limit of the winning solutions.

- All of the QM9 molecules have been optimized to form a local minimum on the potential energy surface (equilibrium). In the future we might be interested in studying non-equilibrium structures, and the solutions might perform less well on these (such as methods ignoring 3D relationships, and that relies entirely on connectivity). All of the top solutions incorporated 3D relationships, and at this time we have no reason to believe that any of them will perform poorly on a non-equilibrium dataset.

- By relying on quantum chemistry calculations that directly relate a molecular 3D structure to the target property, participants could guess the specific methodology and compute the target properties of the test set, thus knowing the correct answer. Due to the computational program not being widely available, but more importantly due to the high computational cost (which directly translates to monetary cost), we expected that it was unlikely that anyone would 'cheat' in this way. This is particularly true as it was a requirement of the competition that the source code to the solution was made accessible to the organizers for a team to be eligible for any prize money.

## S3 Dataset generation

Of the 130,831 molecules in QM9 that passed the geometry consistency check, we removed an additional 42 molecules as these had no hydrogen atoms. The remaining structures were modified to a valid xyz format and input files for Gaussian NMR computations were constructed. The scalar coupling constants (including contributions from separate terms), dipole moment vectors, nuclear shielding tensors, mulliken charges and potential energy were parsed from the Gaussian output files and the dataset was written in an easily accessible csv format. The molecule structures were made available in xyz formatted files as well as a single csv file. While the target observable was the scalar coupling constants, the auxiliary data was included in case additional learning from these could be achieved. A further 14 molecules were removed due to one or more of the scalar couplings being a big outlier compared to the range seen in the remaining molecules. For a small subset of the remaining molecules (108) Gaussian automatically enabled symmetry in the DFT computations. This had little effect on most of the observables, however the direction of the dipole vector and the nuclear shielding tensor became arbitrary as a result of this (however the norm of the vector and trace of the matrix was still correct). This was discovered when the competition was ongoing, but since it only affected a small subset of the auxiliary data, the competition data was not updated with data computed with symmetry disabled.

## S4  Competition details

The goal of the competition was to predict the scalar coupling constants from the interaction between a hydrogen and a hydrogen, carbon or nitrogen atom 1,2 or 3 bonds apart. This was to be done without the competitors knowing any information other than the coordinates and type of element of the atoms in the molecules. The dataset was randomly split into a 65%/35% training/test split. The scalar coupling distributions seemed to be very similar with random splits so stratified splits did not seem necessary.

The competitors were provided the following files:

- *structures.csv*, which contains the coordinates and element type of each of the 2,358,657 atoms of the 130,775 molecules that constitutes the combined test and training set.

- *train.csv*, which contains the values in Hz of the 4,658,147 scalar coupling constants in the training set as well as which atom pairs in which molecule the coupling is between.

- *test.csv*, which contains 2,505,542 pairs of atoms and their respective molecules, which the competitors had to predict the scalar coupling constants of.

- *potential_energy.csv*, which contains the potential energy in Hartree of the 85,003 molecules in the training set.

- *magnetic_shielding_tensors.csv*, which contains the XX, XY, XZ, YX, YY, YZ, ZX, ZY and ZZ components of the magnetic shielding tensors in ppm of the 1,533,537 atoms in the training set.

- *mulliken_charges.csv*, which contains the Mulliken charges in atomic units of the 1,533,537 atoms in the training set.

- *dipole_moment.csv*, which contains the dipole moments in Debye of the 85,003 molecules in the training set.

- *scalar_coupling_contributions.csv*, which contains the FC, SD, PSO and DSO components in Hz of the 4,658,147 scalar coupling constants in the training set. The sum of these equates to the values in *train.csv*.

The test set was further split into a 29%/71% public/private test set. All submissions were scored immediately on both splits. The public score was made available on the public leaderboard, while the private score (which the winners were determined from) were hidden until the end of the competition. here were no changes in the top 37 placements between the public and private leaderboard, indicating that there was little noise in the data.

We opted to use a custom score function to evaluate the submissions. The dataset included 8 different types of coupling: 1JHC (coupling between a hydrogen and a carbon separated by 1 covalent bond), 1JHN, 2JHH, 2JHC, 2JHN, 3JHH, 3JHC and 3JHN. Since the number of couplings of each type differed dramatically (811,999 3JHC couplings in the test set, but only 24,195 1JHN) and spanned different ranges, a bad choice of scoring metric could easily be dominated by the performance on the 3JHC coupling constants.

Our loss function is the logarithm of the geometric mean of the mean absolute error for each type:

$$\text{score} = \frac{1}{T} \sum_{t}^{T} \log \left( \frac{1}{n_t} \sum_{i}^{n_t} |y_i - \hat{y}_i| \right) \tag{S1}$$

Where:

- $T$ is the number of scalar coupling types.
- $n_t$ is the number of observations of type $t$.
- $y_i$ is the actual scalar coupling constant for the observation $i$.
- $\hat{y}_i$ is the predicted scalar coupling constant for the observation $i$.

This way, a 10% improvement in one type of coupling will improve the score by the same amount as a 10% improvement in another type of coupling.

## S5 Analysis

### S5.1 Ensembling

We created an ensemble of the top 400 submissions to see which submissions contributed the most. Of the top 400 submissions, 18 were removed due to being duplicates. The data points (the test set referenced above) were split into a training and test set of equal size, stratified by coupling type using the Scikit-learn library[7]. We did a linear fit to the submissions using TensorFlow[8], minimizing the loss function in equation S1 under the constraint that the weights were positive and summed to 1. 6 submissions had weights of larger than 0.02 as shown in Table S1.

Similarly we did separate linear fits for each coupling type that minimized the mean absolute error. A comparison of the performance of the individual teams and the ensembling strategies described above is shown in Table S2.

Table S1: Competition rank, team name and individual weight of 6 teams whose submission had a weight greater than 0.02 in the ensemble.

| Rank | Name | Weight |
|---|---|---|
| 1 | hybrid | 0.204 |
| 2 | Quantum Uncertainty | 0.270 |
| 3 | [kakr] Solve chem together | 0.111 |
| 4 | Hyperspatial Engineers | 0.203 |
| 5 | DL guys | 0.046 |
| 12 | Team Bird Brain | 0.149 |

Table S2: The performance of the ensemble fitted on all coupling types ('E'), the ensemble fitted on each coupling type separately ('E*'), and the submitted solutions (individual contributions and the total score per coupling type).

| Rank | 1JHC | 1JHN | 2JHH | 2JHC | 2JHN | 3JHH | 3JHC | 3JHN | Total |
|---|---|---|---|---|---|---|---|---|---|
| E | -0.296 | -0.311 | -0.509 | -0.432 | -0.463 | -0.502 | -0.418 | -0.483 | -3.414 |
| E* | -0.297 | -0.332 | -0.515 | -0.433 | -0.463 | -0.507 | -0.421 | -0.485 | -3.453 |
| 1 | -0.284 | -0.288 | -0.477 | -0.410 | -0.440 | -0.477 | -0.400 | -0.464 | -3.241 |
| 2 | -0.277 | -0.291 | -0.458 | -0.410 | -0.450 | -0.474 | -0.394 | -0.472 | -3.226 |
| 3 | -0.269 | -0.244 | -0.480 | -0.411 | -0.443 | -0.488 | -0.397 | -0.461 | -3.193 |
| 4 | -0.267 | -0.286 | -0.482 | -0.401 | -0.433 | -0.470 | -0.390 | -0.455 | -3.184 |
| 5 | -0.275 | -0.289 | -0.468 | -0.400 | -0.428 | -0.452 | -0.387 | -0.450 | -3.149 |
| 12 | -0.221 | -0.278 | -0.448 | -0.357 | -0.414 | -0.444 | -0.336 | -0.437 | -2.936 |

### S5.2 Correlation

We investigated how similar the submissions from the top teams were by computing the correlation between a scaled subset of their submissions on the test data. 20,000 data points for each coupling type were drawn randomly from the test set. For each coupling type the submissions were the scaled by their individual root mean square error (RMSE). These two pre-processing steps were done to make each coupling type have an equal impact on the correlation.

The same analysis was done on the top 50 team submissions where the submissions were clustered with hierarchical clustering using the 'complete' linkage in the Scipy module[9].

### S5.3 Manifold

In a similar analysis, we projected the top 100 submissions of a scaled subset of the test set down on a two-dimensional manifold. Again, 20,000 data points for each coupling type were drawn randomly from the test set. As the mean absolute error (MAE) is a more natural metric in relation to the scoring metric used in the competition, the submissions were for each coupling type scaled by their individual MAE. This pre-processing is important as if the submissions weren't scaled appropriately, the dimensionality reduction might mainly capture the average performance of a given submission, rather than how the methods differ in general.

Figure S1 shows the submissions projected onto a two-dimensional Multidimensional Scaling (MDS) manifold (which tries to preserve the Manhattan distance between submissions)[7, 10, 11, 12].

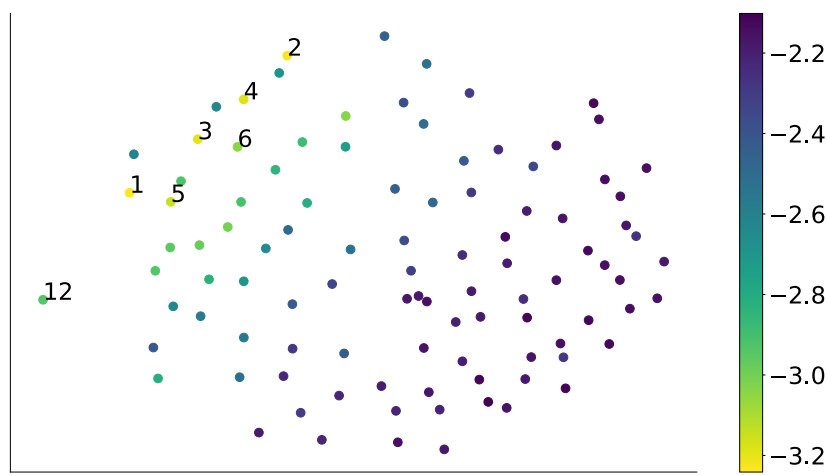

Figure S1: Projections of top 100 submissions to a two-dimensional MDS manifold. Coloring indicate submissions score as indicated by the colorbar. A subset of 7 submissions are marked with the competition rank.

### S5.4 Comparison with previous competitions

To get an idea of how engaging the competition was to the community (and earlier how much participation we could expect), we looked at how the number of participating teams have historically depended on the prize pool. We retrieved the number of participating teams for all previous competitions with a monetary prize and converted the prize into US dollars. Additionally we restricted ourselves to competitions where everyone could enter, where the number of teams were listed, that were not of recruitment or code-type, and competitions that took place within the last five years. Figure S2 shows how this competition compare to previous ones.

### S5.5 Participant engagement

We looked at how many days separated each participating teams' first and last submission, to get an overview over the average engagement. Figure S3 shows

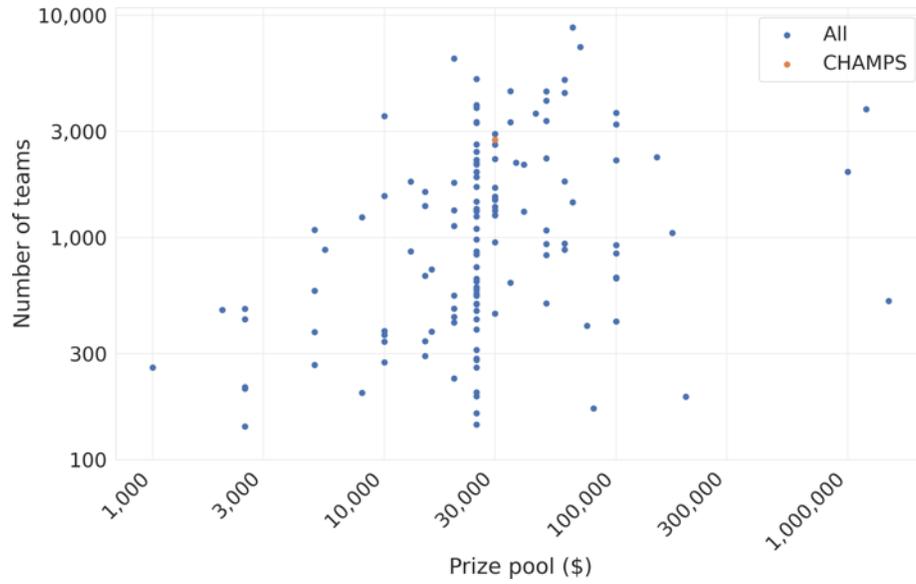

Figure S2: How the prize pool are related to number of participating teams on Kaggle in the last five years (log-log plot). This competition highlighted in orange.

that many teams concentrated their efforts (or did only a single submission) over 1-2 days. However, Figure S4 truncates any team that made continuous submissions over a period longer than 3 weeks into a single bin, showing that the majority of the teams were engaged throughout the competition

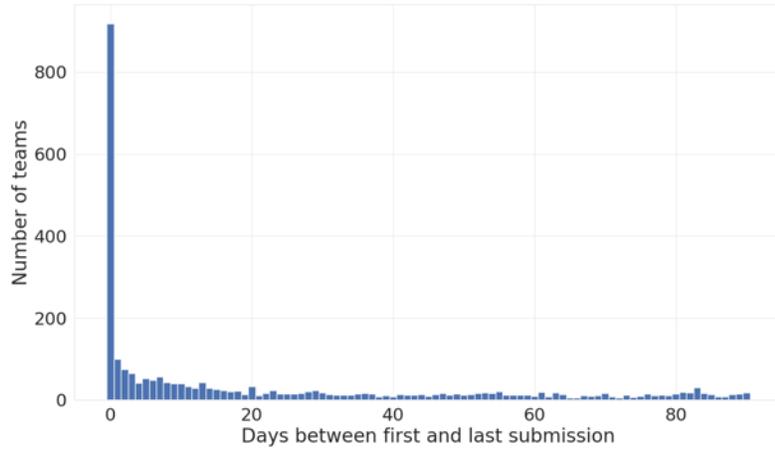

Figure S3: Histogram showing the number of days between a teams' first and last submission.

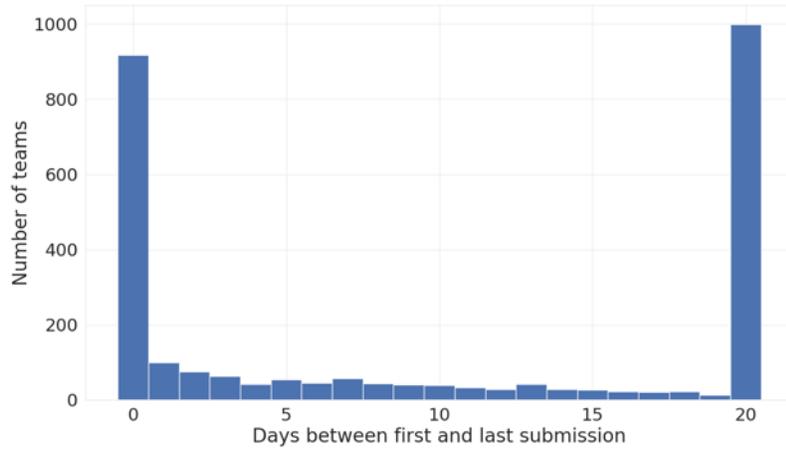

Figure S4: Histogram showing the number of days between a teams' first and last submission, where the rightmost bin indicate teams with more than 3 weeks between first and last submissions.

# S6 Comparison to IMPRESSION

In the manuscript we provide comparisons to our previously published methodology (IMPRESSION [13]). We report ratio's in a range, where the upper bound are the best models that we have trained on the training set. These could likely be improved upon using higher memory machines.

As the error in kernel methods tend to scale as a power law with number of data points, we extrapolated what the error theoretically could be in the limit of using the entire training data set. This estimate is used as the lower bound. Table S3 shows the score contribution of each coupling type for the best ensemble, the winning submission as well as scores for IMPRESSION best trained models and interpolated errors. Similarly Table S4 shows the mean absolute errors.

Table S3: Score contributions for each coupling type for the best ensemble, the winner and IMPRESSION lower and upper bounds. (lower, upper)

| Type | Ensemble* | Winner | IMPRESSION |
|---|---|---|---|
| 1JHC | -0.296 | -0.284 | (-0.150, -0.003) |
| 1JHN | -0.332 | -0.288 | (-0.256, -0.178) |
| 2JHH | -0.515 | -0.477 | (-0.209, -0.119) |
| 2JHC | -0.433 | -0.410 | (-0.238, -0.018) |
| 2JHN | -0.463 | -0.440 | (-0.283, -0.143) |
| 3JHH | -0.507 | -0.477 | (-0.048, 0.019) |
| 3JHC | -0.421 | -0.400 | (-0.123, 0.049) |
| 3JHN | -0.485 | -0.464 | (-0.216, -0.107) |
| Sum | -3.453 | -3.241 | (-1.522, -0.499) |

Table S4: Mean absolute error for each coupling type for the best ensemble, the winner and IMPRESSION lower and upper bounds.

| Type | Ensemble* | Winner | IMPRESSION |
|---|---|---|---|
| 1JHC | 0.0937 | 0.1031 | (0.3002, 0.9728) |
| 1JHN | 0.0702 | 0.0999 | (0.1292, 0.2404) |
| 2JHH | 0.0162 | 0.0220 | (0.1886, 0.3872) |
| 2JHC | 0.0313 | 0.0376 | (0.1488, 0.8682) |
| 2JHN | 0.0246 | 0.0296 | (0.1040, 0.3180) |
| 3JHH | 0.0173 | 0.0220 | (0.6815, 1.1675) |
| 3JHC | 0.0345 | 0.0408 | (0.3751, 1.4811) |
| 3JHN | 0.0207 | 0.0244 | (0.1780, 0.4257) |
| Geometric mean | 0.0317 | 0.0391 | (0.2183, 0.6069) |

# S7  Solution 1 - Hybrid

## S7.1  Features

Of the provided data, we use only the element and position of each atom and the scalar coupling type of each bond. From these, we use the open source package RDKit[14] to generate additional features for each atom, specifically bond order and partial charge. Below we list the total set of features of each molecular object that we use in the network.

- Atoms: element, number of neighbors, order of neighboring bonds, partial charge, angle with nearest neighbors. All atoms are included.

- Bonds: each atom's element, coupling type (if applicable), bond order (if applicable), distance. These are included regardless of whether there is a chemical bond or not.

- Triplets: each atom's element, bond angle. These are included only if there is a chemical bond (one central atom bonded to a pair of other atoms).

The methods described can be generalized to include quadruplets (and dihedral angles) as well, although the addition of quadruplets was found to not be helpful in this problem.

### S7.1.1  Molecular Representation

Deep learning approaches to problems in molecular modeling generally represent the molecule as a graph, with nodes representing atoms and edges representing bonds. Many of these are then passed through networks that can be described using the general framework of message passing neural networks (MPNN)[15], where information is usually passed through alternating convolutions, first from edges to nodes and then from nodes to edges. This representation is restrictive for several reasons. Most importantly, information may only be passed locally, as direct connections do not exist between, for example, pairs of atoms, or an atom and a bond far away in the molecule. Further, this molecular representation only allows us to use features that directly correspond to a particular atom or bond; for example, it is not clear how to incorporate features of atom triplets as input in a typical MPNN. In contrast to this framework, we represent each molecular object of interest (in this case, each atom, bond, and triplet, but we may incorporate other objects or properties of the molecule as we like) as a node in a complete graph.

## S7.2  Model Architecture

Our architecture is a deep learning approach with three stages: an embedding layer, several graph transformer layers, and element-wise group convolutions. Each of these is detailed in the subsections below.

### S7.2.1 Embedding

We use three different embedding layers, each corresponding to a type of molecular object (atom, bond, and triplet). All of these objects have a mixture of discrete and continuous features, and some of these features are hierarchical in nature (for example, bonds' coupling type and sub-type, which includes some local bond-order information). As such, we use hierarchical embeddings[16], which embed each feature separately, and then linear combine these embeddings to yield a single representation of the object. Discrete features are embedded in the usual fashion, while for continuous features we use a sine filter embedding similar to the position embeddings employed by Transformer models[17]. We embed all features described in Section S7.1, with the notable exception of atom positions. Instead, we use these positions to construct a matrix of relative distances to be used later in the network; this ensures that network output is invariant to translation and rotation of the molecule.

Subsequent layers in the network will convolve over all nodes in the graph; as such, we require that the output size of each of these embeddings be the same. It is worthwhile to note that after the embedding layer, all nodes are treated identically by the network – that is, the graph transformer layer makes no distinction between nodes represented by atoms, bonds, and triplets. Rather, all representation of molecular objects exist in the same space, and it is the job of the embedding layer parameters to learn representations that can be meaningfully distinguished by the remainder of the network.

### S7.2.2 Graph Transformer Layers

Throughout this section, we let Z be a matrix that represents the hidden layer of a molecule, where each column of Z represents a particular node. The layer describe here, which we call the graph transformer layer, forms the foundational building block of our architecture. This layer is an extension of a graph convolution layer, which, in its most general form, may be represented as

$$\mathbf{GraphConvolution}(Z) = \sigma(WZA) \quad \text{(S2)}$$

where $\sigma$ is an activation function, W is a matrix of learnable weights, and A is a static mixing matrix that encodes the structure of the graph; common choices for A include a normalized adjacency matrix or a graph Laplacian. Our first modification is the replacement of the mixing matrix A with a self-attention mechanism, which was popularized by the Transformer model[17] for sequence modeling. In self-attention, the strength of message passing is computed via a parametrized inner product between nodes:

$$\mathbf{GraphSelfAttention} = \sigma(W_3 Z \quad \mathbf{softmax}(Z^T W_1^T W_2 Z)) \quad \text{(S3)}$$

Under this architecture, W3 learns the message to be passed, while the weight matrices W1 and W2 learn the strength of each message. Replacing the mixing matrix A in this manner removes the only structural information the network receives about the graph. To reintroduce this information, we incorporate a scaling based on the distance between each pair of nodes in the graph. Let D be a matrix such that $D_{i,j}$ represents distance between nodes i and j, and let $\gamma$ be a learnable scalar parameter. Distance-scaled self- attention is given by

$$\textbf{ScaledGraphAttention}(Z) = \sigma(W_3 Z \quad \textbf{softmax}(Z^T W_1^T W_2 Z - \gamma D)) \quad \text{(S4)}$$

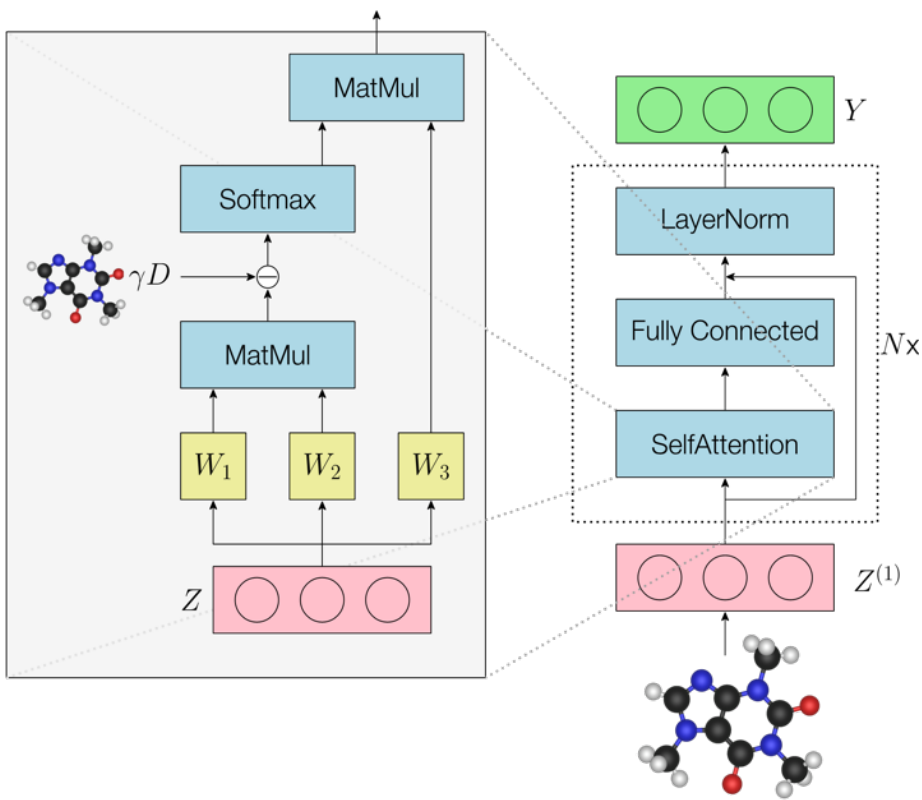

Figure S5: Graphical representation of the model architecture.

This scaling has the effect of reducing the strength of interaction between pairs of faraway nodes. The learnable parameter $\gamma$ allows us to learn the intensity of this scaling: as $\gamma$ approaches 0, relative distances are ignored, and as $\gamma$ approaches $\infty$, we operate on a sparse graph, with information only flowing between directly adjacent molecular objects. The scaled graph self-attention transformation is shown in Figure S5. As in the original Transformer model, we follow self-attention with a linear transformation, a residual connection, and layer normalization[18]. The full graph Transformer layer is given by

$$\text{GraphTransformerLayer}(Z) \tag{S5}$$
$$= \text{LayerNorm}(W_4 \text{ScaledGraphSelfAttention}(Z) + Z) \tag{S6}$$
$$= \text{LayerNorm}(W_4 \sigma(W_3 Z \text{softmax}(Z^T W_1^T W_2 Z - \gamma D)) + Z) \tag{S7}$$

### S7.2.3 Defining Distances

The self-attention scaling described above requires a matrix D of relative distances between objects represented in the graph. However, since bonds and triplets do not have a well-defined position in space, it is not necessarily trivial to define distances among these objects. To do so, we begin with distances between two atoms a and a', denoted $d_{atom}(a, a')$, which we define as simple Euclidean distance between their positions. Distance involving larger sets of atoms are defined recursively, and with the guiding principle that the distance between two sets of atoms should be 0 if and only if one set is contained in the other. Distances are defined as below. In the following, let a be an atom, B be a bond connecting atoms b1 and b2, and C be a triplet containing center atom c1, other atoms c2 and c3, and bonds c12 and c13 .

$$d_{\text{atom,bond}}(a, B) = \min(d_{\text{atom}}(a, b1), d_{\text{atom}}(a, b2)) \tag{S8}$$

$$d_{\text{bond,bond}}(B, B') = \frac{1}{4} \sum_{b \in B, b' \in B'} d_{\text{atom,bond}}(b, b') \tag{S9}$$

$$d_{\text{atom,trip}}(a, C) = \min_{c \in c_1, c_2, c_3}(d_{\text{atom,atom}}(a, c)) + d_{\text{atom,atom}}(a, c_1) \tag{S10}$$

$$d_{\text{bond,trip}}(B, C) = \min(d_{\text{bond,bond}}(B, c_{12}), d_{\text{bond,bond}}(B, c_{12})) \tag{S11}$$

$$d_{\text{trip,trip}}(C, C') = \frac{1}{4} \sum_{c \in c_{12}, c_{13}, c' \in c'_{12}, c'_{13}} d_{\text{bond,bond}}(c, c') \tag{S12}$$

### S7.2.4 Group Convolutions

After being passed through some number of graph transformer layers, we take the nodes representing coupling bonds and pass them through a series of $1 \times 1$ convolutions inside of a residual block. These convolutions are grouped according to coupling type and bond order, such that the set of convolutional filters applied is different for each group. A final grouped convolution outputs the predicted scalar coupling constant.

## S7.3 Training and Ensembling

In total, 13 of the models described in the previous section were trained on the task of predicting magnetic scalar coupling constants. Model size varied, but generally had between 12 and 18 graph transformer layers, and hidden dimension between 600 and 800. These models were initially trained using absolute loss instead of the loss defined in equation S1. The model was implemented in PyTorch[19] and network training was done using the ADAM optimizer[20] with a cosine annealing learning rate scheduler[21] for 200 epochs. For some of our best models, we additionally fine-tuned by training for approximately 40 epochs using the loss defined in equation S1 to improve their single model scores by around -0.020 to -0.030. For each coupling subtype (i.e. coupling type, plus further breakdowns by element and chemical environment information), the targets were scaled to 0 mean and one standard deviation to simplify the learning process. These 13 models were ensembled with a mix of mean and median ensembling. First, for each coupling type, we observed how frequently each model's prediction was the median among all 13 predictions. The 9 best models according to this metric were selected, with the remaining 4 discarded. Finally, for each coupling constant, we selected the center 5 median values from among the 9 model predictions; the mean of these 5 values was used as the final prediction. We found that this ensembling strategy was more effective than strict mean or median ensembling, and did not require a validation set to estimate individual models' test accuracy.

## S8 Solution 2 - Quantum Uncertainty

### S8.1 Data Augmentation

The key to the success of this model lies in the data augmentation, and comes from the insight that pair-wise properties can be modelled as atom-wise properties, given that relative position to the paired atom is encoded in the input. This is achieved by making duplicate 'sibling' molecules that are translated to have a different atom as origin.

For a given molecule of $N$ atoms, where $M$ of these are of atom type H, C or N, $M$ siblings are constructed, each centered on a different H, C or N atom. Furthermore the sibling molecules are padded with dummy atoms to make all molecules have the same number of atoms (29 for this competition), which are later masked in the model. Each atom in the sibling molecules are then matched with the corresponding coupling constant label, where dummy values are used for coupling types that are not present in the dataset, and these are similarly masked later in the model.

### S8.2 Input Features

A feature vector is constructed for each atom in the sibling molecules, that contain the Cartesian coordinate of the atom, the atom type index (C=0, H=1, ...) as well as an index for the type of coupling formed with the origin atom (1JCH=0, 2JHH=1, ...). In the feature vector 3 elements will be the X, Y and Z component of the Cartesian coordinate, while the remaining elements are used for embedding the atom type and coupling type. For a feature vector of e.g. length 1024, 510 elements will be repetitions of the atom type index, 511 will be repetitions of the coupling type index, while the remaining 3 elements will contain the Cartesian coordinates. Note that there is no graph information nor any other manually engineered features.

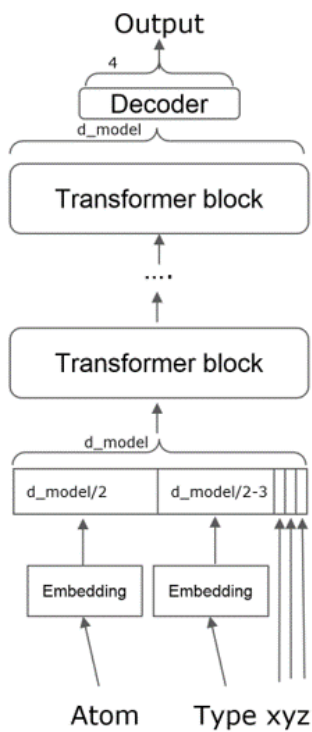

Figure S6: Overview of the overall architecture

### S8.3 Model Architecture

The model used a standard transformer architecture [22] utilizing the fast.ai library [23], where each feature vector are updated with several transformer layers followed by a feed forward neural network to decode the scalar coupling constants from the transformed feature vectors (See Fig S6). Adding rotations during training didn't improve the model performance, indicating that the molecules were either systematically aligned in a way that enabled efficient training, or that rotation invariance were easily learned by the model.

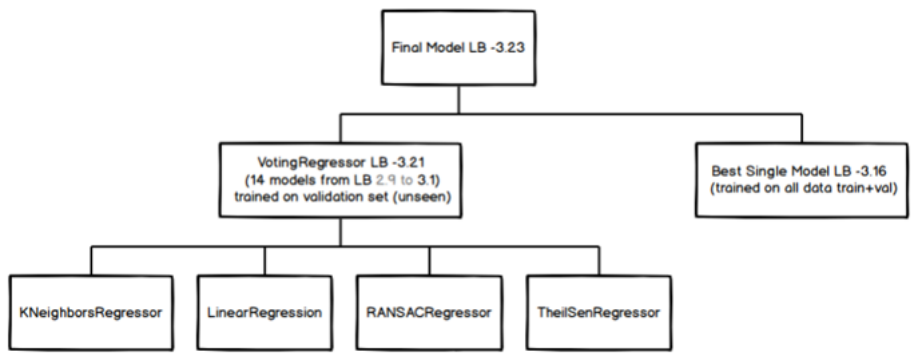

Figure S7: Overview of the ensembling strategy

### S8.4 Training and Ensembling

14 models were trained, with varying dimensions from 512 to 2048 and layers from 6 to 24, and with scores of -2.9 to -3.1 Each model parameter size ranged from 12M to 100M (biggest model). A small validation set were kept separate to fit the ensemble model. Four different regressors were trained on the validation set predictions of the 14 models: k-nearest neighbors, linear, RANSAC [24] and Theil-Sen [25] regression models. A voting regressor were using to combine predictions of the four models, which resulted in a score of -3.21. Finally the single model with the best score (6 layers, 1024 dimensional feature vector) were trained on all the training data (including validation set), yielding a score of -3.16. Averaging this best single model with the voting regressor model yielded a final score of -3.23 on the public testing set. An overview of the ensembling strategy is shown in Fig S7. Predictions from the single best model takes ∼10 minutes for the test set of 46K molecules (∼15 ms per molecule).

## S9 Solution 3

### S9.1 Features

Due to the graphical nature of molecular systems, graph-based architectures are often applied to many chemical problems including this competition [26]. However, the nature of problem in this competition is slightly different to typical chemical problems, since atom-pair properties of three-dimensional chemical structures has not been modelled before and only a subset of atomic pairs in the molecules have target values (coupling constants). For example, the largest molecule in the training set (nonane) has 29 atoms and there are 406 atomic pairs. Among those atomic pairs, only 127 atomic pairs have target values. For this reason, an alternative representation for molecular systems than the graph representation can be utilized. Here, we use a set of only a part of pair features as a descriptor for a molecule. Even though we replace the graphical representation, the size-extensiveness and permutation-invariant nature should be preserved. A similarity-based attention (or simply attention) is one of the mechanisms that satisfies both conditions:

$$\text{Attention}(Q, K, V) = \text{softmax}(\frac{QK^T}{\sqrt{d_k}})V \tag{S13}$$

where Q,K,V and $d_k$ are query, key, value, and feature dimension of the key value respectively. An output of attention is a weighted average of input values where the weights are determined by similarity of queries and keys. Attention layers hold permutational invariance and size-extensivity because if the size or order of the input vector is changed, those of corresponding key and query vectors are changed. In many different contexts of machine learning, a large number of variants of attention have been proposed. A multi-head attention is one of the most frequently employed ones. It generalizes conventional attentions with concatenating results of attention whose inputs are linearly transformed with different weights.

$$\text{Multihead}(Q, K, V) = \text{Concat}(O_1, O_2, ..., O_h)W^0 \tag{S14}$$
$$O_i = \text{Attention}(QW_i^Q, KW_i^K, VW_i^V) \tag{S15}$$

where $W_i^Q$, $W_i^K$, $W_i^V$ and $W^O$ are weights for query, key, value, and outputs of attentions respectively. One of the benefits of using (multi-head) attention is to minimize sequential computing which is an obstacle for parallelization.

## S9.2 Model Architecture

Transformers, one of the well-known natural language processing architectures, recorded overwhelming performance in the Seq2Seq[27] problem with replacing sequential computing to multi-head attention. The original transformer model[22] includes positional embedding to impose sequential information on input and output feature vectors before performing the encoding and decoding process, since encoder and decoder are composed of permutation-invariant layers. Also, like ordinary Attention and Linear blocks, they are size extensive which means different lengths of sequence can be treated. Hence, we employ encoder of Transformer to embedding pair sequence information to construct the latent space that contains interaction among atomic pairs. In terms of NLP, atomic pair information and pair sequences become words and sentences respectively. We refer to the original transformer paper for more details on the attention and transformer architecture [22]. Once again important features of the model is size-extensivity and permutation-invariance.

Our model (shown in Figure S9) adopts the encoder of Transformer architecture because of its size extensivity and permutation invariance. We make an atomic pair sequence to represent a molecule and calculate a target value by considering interactions among atomic pairs. The Transformer encoder (grey part in Figure S9) is employed to transfer input feature to latent space which is read by Readout (purple part in Figure S9, more detailed information in Figure S10). For each type, weights of Readout are different but structures and dimensions of weights are the same. (see Table S8)

| Encoding | Number of encoder layers | 8 |
| --- | --- | --- |
| | Number of heads for attention | 8 |
| | Dropout Ratio | 0.1 |
| | Intermediate size | 2048 |
| Readout | Dimension | 832 |
| | Dropout Ratio | 0.1 |
| Embedding | Embedding Dimension for Atomic Charge | 32 |
| | Embedding Dimension for Position | 256 |
| | Embedding Dimension for Atomic Number | 64 |
| | Embedding Dimension for Distance | 64 |
| | Embedding Dimension for Type | 64 |
| Data Augmentation | Mean of translational noise (Angstrom) | 0 |
| | Standard deviation of translational noise (Angstrom) | 2 |
| | Mean of angle in rotational noise (rad) | 0 |
| | Standard deviation of angle in rotational noise (rad) | 1.57 |
| Dummy Types | 1JHO, 1JCO, 1JCN, 1JNO, 1JCC, 1JNN, 1JFC | |
| Training and Ensemble | Learning rate | 0.0003 |
| | Linear warmup step | 30 |
| | Weight Decay | 0.01 |
| | Size of seed ensemble | 10 |
| | Mean of initial weights distribution | 0 |
| | Standard deviation of initial weights distribution | 0.02 |

Figure S8: Detailed solution information

### S9.3 Readout stage

Readout stage evaluates the spin coupling constant (SC in Figure S9) from the output sequence, output of the encoding layer. Here, we employ physical condition; the spin coupling constant is the sum of 4 components, Fermi Contact contribution (FC), Spin-dipolar contribution (SD), Paramagnetic spin-orbit contribution (PSO) and Diamagnetic spin-orbit contribution (DSO). The Readout stage is designed to predict a target value as mean of two differently evaluated values. One directly comes from the latent space and the other one is a sum of 4 components which comes from latent space. The loss function for the whole

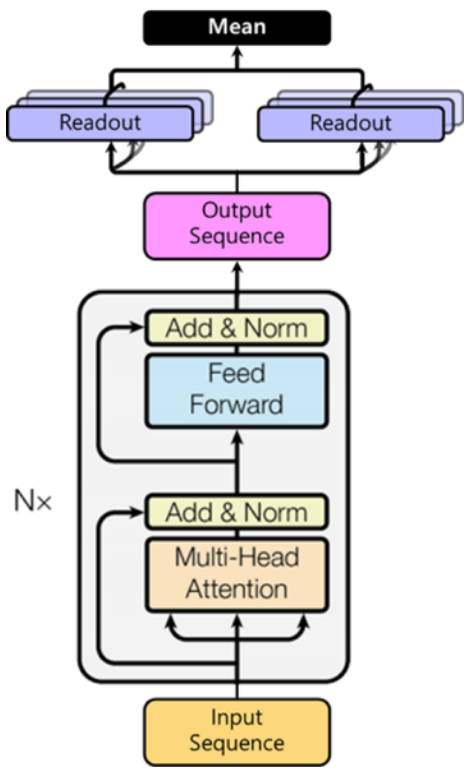

Figure S9: Pictoral representation of the overall architecture

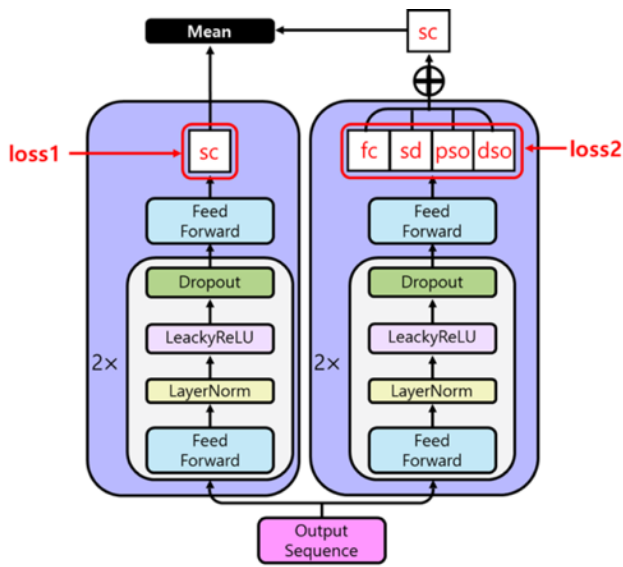

Figure S10: Pictoral representation of the overall model

model is constructed as the log of the sum of MAEs of outputs of two Readout layers.

### S9.4 Features

An atomic pair feature is composed of two atoms' properties and pair properties. (See Figure S11 ) For atomic properties, the position of atom, atomic number and atomic charge, available in QM9 dataset are used and distance between them and type of coupling are for the pair properties. Those values are embedded and concatenated to build feature vector. Unlike the sum operation, concatenation is not a commutative operation so our feature vectors are dependent on the order of atoms in a pair. To make the feature vector independent of the order of atoms, test time augmentation or data augmentation can be applied, but in the data set, the order of atoms in a pair are sorted. Therefore, we use none of them. For clarity, it should be noticed that this order-dependency does not necessarily lead to permutation dependency of atomic pairs belonging to molecules.

#### S9.4.1 Data Augmentation

As described in the previous section, positions of atoms are used in input sequences, therefore rotational and translational invariances are not conserved in our model. Although both invariances are not mathematically preserved, by training augmented data, we make parameters of our model guided to have pseudo-invariance in a certain range of translational and rotational changes.

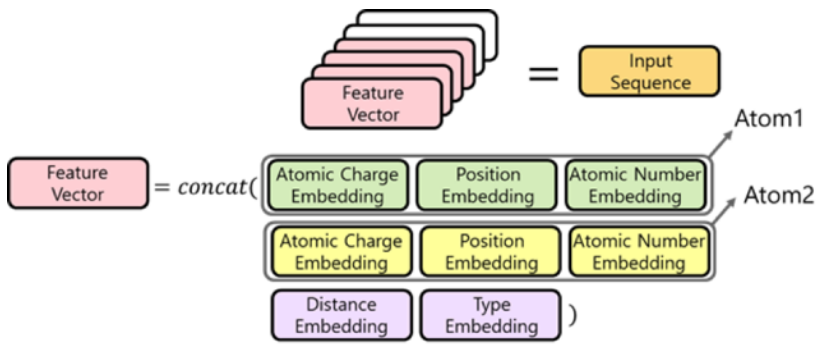

Figure S11: Pictoral representation of the input features

Fortunately, positions of atoms belonging to the QM9 set do not have extreme values, which means this kind of less rigorous strategy can cover all the test cases in the competition. If we need to cover general molecular geometry, this kind of strategy might not work. Generation of noise can be done in every epoch with low computational cost, the size of the original training database being enlarged 10 times with randomly selected noise. The selected noise is not changed during the training. The translational perturbations on each axis and angle of rotational perturbations are sampled from a Normal distribution. By this augmentation of the training data, we observed that the trained model recorded small enough disagreements with both changes and a better score in leaderboard. Mean and standard deviation values for both translational and rotational perturbation can be found in Table S8.

### S9.4.2 Dummy types

Some of the geometric information of molecule is missing in the input sequences because not all atomic pairs are included. It may cause an incomplete description of chemical circumstances. In order to overcome this limit, dummy pairs are introduced. The dummy pairs are a part of the input sequences but do not have coupling constant values. The predicted values of dummy types are not used to evaluate the loss but the feature vectors of dummy types participate in process of building latent space even for meaningful pairs. By this, the geometric information of non-activated pairs can be included in the input sequence. Thousands of pair types exists in the given training set. Among them, 7 additional atomic types are included in input sequences: 1JHO, 1JCO, 1JCN, 1JNO, 1JCC, 1JNN, and 1JFC. The selection of these dummy pairs is a tunable hyperparameter but due to a lack of computing power, we did not test various combinations of dummy types. We choose to add neighboring types first because nearby pairs may have a strong impact on chemical circumstances. If more dummy types are added, more geometric information can be included in input sequences but an increase of computational costs follows as the length of sequence increases.

## S9.5 Training and Ensembling

In order to achieve a high score, we employed an ensemble technique and pseudo-labeling. An ensemble technique is to merge the results of various models and pseudo-labeling is to use new dataset which contains unlabeled data whose label is assumed as the results of previously trained model. We originally planned to use a seed ensemble which uses a number of identical models with different random seed numbers but because of inequalities of teammates' computing resources. Therefore, models with different N values (in Figure S9) and epochs are used. An overall training process is illustrated in Figure S12. At first stage, 15 models are trained using only training dataset with dropout. At second stage, parameters of models were fine-tuned by turning off the dropouts. By using models trained up to the second phase, labels for test sets (so-called pseudo-label) are predicted. To minimize error of pseudo-label, results of 4-8 trained models up to the second phase are employed. In the third phase, newly constructed training sets which including both original training dataset and pseudo-labeled test set are used. Because of due date of competition, we only take 8 models and further trained with the expanded datasets. At phase 3 and 4, 20 epochs are progressed. To mitigate the unpredicted bias from pseudo-labeling, at the last phase, the models are trained with only the original training dataset. Adam optimizer, typical choice, is employed with gradient clipping. Linear warmup and linear decay are used for the stable convergence of training. All weights are initialized with uniform distribution (mean=0.0, std.=0.02) and 0 is used for initial bias values.

### S9.5.1 Specification of models

Training in each ensemble is composed of 4 different steps. First step, training is performed with training set data and dropout. In the second step, further training is proceeded without dropout. Using the obtained model from the second step, the coupling constants for testing set are predicted; these predicted values are called pseudo label. The pseudo-labeled data in addition to training data is used in the third training step. To mitigate overfitting, further training is performed with only training set.

Output values of each type are normalized because each type shows different distribution of target values.

## S9.6 Performance

Our final model has around 75M parameters. With two V100 graphic cards, training the model takes around 2 days.

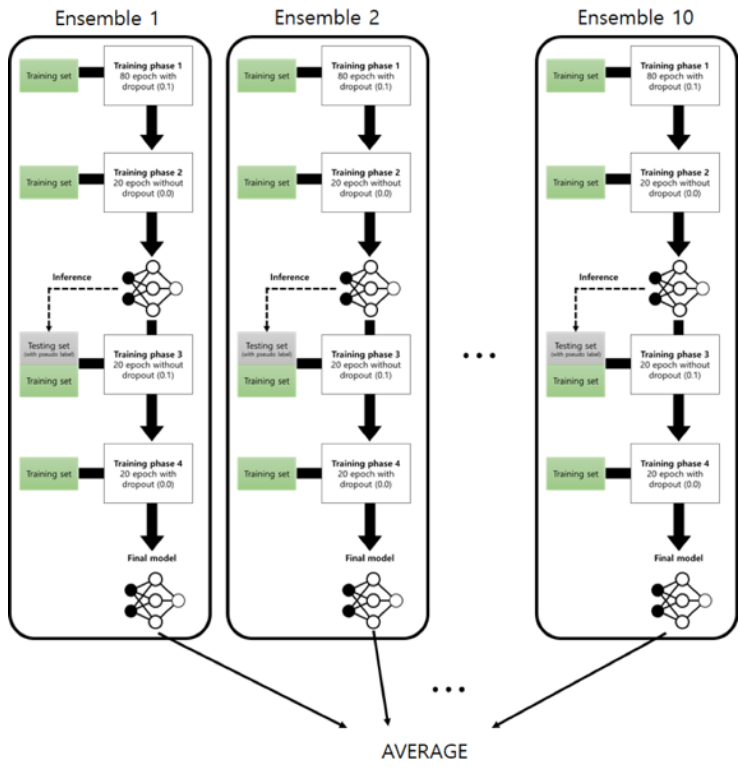

Figure S12: Pictoral representation of the training procedure

# S10 Solution 4

## S10.1 Model Architecture

The model we used is a development of the Graph Attention Network [28] architecture with Gated Residual Connections [29]. The model operates on molecular graphs (nodes being the atoms, and edges being the bonds), augmented by artificial links between the nodes representing atoms 2 and 3 bonds apart in the molecular structure. Both graph nodes and graph edges hold feature vectors (in contrast to the more standard approach where only the nodes hold feature vectors). Size of the node and edge feature vectors are equal in each layer. The network is implemented in the PyTorch [30] framework, using the Deep Graph Library [31], which requires all graph edges to be directed. We therefore represent each bond (as well as each 1-3 and 1-4 link) as a pair of graph edges (one in each direction).

The network consists of multiple layers, where each layer updates both the node and the edge feature vectors through the following steps:

1. Fully connected layers applied to nodes and edges:

$$\eta^l = W_{node}^l n^{l-1}, \epsilon^l = W_{edge}^l e^{l-1}$$

   where $n^{l-1}$ and $e^{l-1}$ represent the values of the node and edge embeddings from the previous layer, while $\eta$ and $\epsilon$ represent intermediate values used within the rest of the layer.

2. Attention based convolutional update applied to each node:

$$\alpha_{ij}^l = softmax_j(\sigma(A^l[\eta_i^l \parallel \epsilon_{ij}^l \parallel \eta_j^l]))$$

$$n_i^l = \sigma(\sum_j (\alpha_{ij}^l \eta_i^l \epsilon_{ij}^l))$$

   where $\sigma$ is a non-linearity (in the final model we used the Leaky Rectified Linear Unit with the slope set to 0.2). Furthermore, we employ the multi-head extension to the attention mechanism (as described in [28]), where multiple independent attention mechanisms are applied, and their results concatenated:

$$n_i^l = \bigparallel_{k=1}^{K} \sigma(\sum_j (\alpha_{ijk}^l \eta_{ik}^l) \epsilon_{ijk}^l)$$

3. Fully connected update to each edge, based on the concatenation of the source node feature vector, edge feature vector, and the destination node feature vector:

$$e_{ij}^l = W_{triplet}^l[n_i^l \parallel \epsilon_{ij}^l \parallel n_j^l]))$$

4. At the end of each layer we apply layer normalization [32] to node and edge feature vectors.

Each layer is followed by a Gated (Parametric) Residual connection [29], on both the node and the edge feature vectors. The outputs of the residual values are passed through Parametric Rectified Linear Units [33].

The final model is made up of an embedding layer, eight residual layers, and the two fully connected output layers applied to edge feature vectors of the final layer. Embedding size for both the nodes and the edges in each residual layer is 1152, and the number of heads is 24 (each head accounting for 48 parameters). The output sizes of the two final fully connected output layers were 512 and 8 (one output for each of the coupling types).

Coupling constant between two atoms are predicted by taking the corresponding output from the edges linking the graph nodes representing the two atoms. For each pair of atoms the graph will incorporate two such edges (one in each direction, due to the directed nature of the graph). During training, we treat these as two independent predictions (calculating and back-propagating the loss as if these were two data points). In prediction mode we output the average of these two predictions.

### S10.2  Features

In addition to the molecular graph structure (atoms and bonds as inferred by OpenBabel [34]) the model uses a number of atom and bond features. For atoms we use: electronegativity, first ionization energy, and electron affinity for each atom type, as well as atom Mulliken charge taken from the QM9 dataset [35, 36]. For edges we use features: bond length, bond angle (for artificial 1-3 edges), and the dihedral angle (for artificial 1-4 edges). All features were standardized based on the training set statistics. Labels were standardised for all models, except those used for predicting 1JHC, where we only subtract the mean (which we empirically determined to produce better results).

### S10.3  Training and ensembling

We trained the model using a modified version of the LAMB optimizer [37], where we decouple the weight decay term from the trust region calculations, similarly to the AdamW modification of the Adam optimizer [38]. The training was done using a mini-batch size of 80 molecules. Models were first pre-trained to jointly predict the scalar coupling constants for all coupling types. In the fine-tuning step we train separate models for JHC and JHH types, and continue training the joint model for the JHN types. The training was done using the Mean Absolute Error loss.

During training we used the following dynamic learning rate schedule:

- 30 epoch cycle, linearly varying the learning rate from 0.001 to 0.01 and back (pre-train phase)

- 70 epochs with a constant learning rate of 0.001 (pre-train phase)

- Constant learning rate of 0.001 until convergence; 90-100 cycles, depending on coupling type (fine tuning phase)

The model was regularized using weight decay set to 0.05 for the first 30 epochs and 0.01 afterwards. Final model parameters are derived by running Stochastic Weight Averaging [39] over the models from the final 25 epochs of training. Final predictions are produced as the mean of the outputs from two folds (two sets of models, both using the same architecture, but with a different train/validation split, with 90% of the data used in training, and 10% used for validation). Each of the two model sets consists of 6 models:

- a model per coupling type, for each of 1JHC, 2JHC, 3JHC, 2JHH, 3JHH coupling types

- 1 model specialized for 1JHN, 2JHN, and 3JHN coupling types

### S10.4  Performance

The training procedure for the final ensemble took 200 GPU hours on systems with 2080Ti cards. The ensemble achieved a score of -3.18667 on the public test set, and -3.18085 on the private test set provided by the competition. In order to illuminate the architecture's performance, in addition to the final model (denoted FULL) we also benchmarked a single model trained to predict all of the coupling type interaction jointly (denoted SINGLE), and an ensemble where single model is specialized for each coupling type (denoted TYPE). The results are presented in Table S5.

Table S5: Scores achieved by different ensembling choices.

| Model | Private score | Public score | Train time (GPU h) |
|---|---|---|---|
| FULL | -3.18085 | -3.18667 | 200 |
| PER-TYPE | -3.13853 | -3.14362 | 85 |
| SINGLE | -2.96183 | -2.96443 | 24 |

## S11  Solution 5

Our model is an ensemble model built from 16 graph-based deep learning models. Our base models are inspired from MatErials Graph Network (MEGNet)[40] which is an architecture used to predict properties of either molecules or crystals. In the following we will describe our best single model.

### S11.1  Input features

We created features from the raw atom elements and coordinates using OpenBabel[34]. OpenBabel provides numerous chemical features for each atoms, bonds, and for the whole molecule, which are all included in our input features. We also added translation and rotation invariant features such as ACSFs[41], distances of bonds, angle between bonds and the raw coordinates. Random rotations on the molecule coordinates were applied as a way of data augmentation. The most important features sorted by a permutation feature importance are the following:

- **Atom**: size of smallest ring that contains the atom, heteroatom valence, number of ring connection to the atom, average angle of bonds connected to the atom, whether a neighboring atom has an unsaturated bond to a third atom, count of hydrogen bound to the atom.

- **Bond**: angle formed by the neighboring bond with the closest atom, minimum distance of neighboring bond, bond distance, scalar coupling type.

- **Molecule**: number of atoms, number of non-hydrogen atoms.

We also noticed that the most important features of our preliminary models, according to a permutation feature importance, were the ring topology in the molecule and the angles between two bonds. This inspired us to modify MEG-Net by adding two operations related to rings and bond-bond angles. We will describe these modifications below.

## S11.2 Model architecture

Our models consists of three stages :

1. a preprocessing operation that transforms molecular, bonds and atomic features into a graph representation.

2. multiple steps of a graph update operation.

3. a readout operation that transform the graph representation to a set of scalar coupling values.

The architecture is described in Figure S13.

### S11.2.1 Preprocessing operation

The preprocessing operation builds a graph representation of the molecule. Each atom in the molecule represents a node in the graph. We choose to represent the molecule with a dense graph rather that limiting the edges to chemical bonds. This choice helped the information flows better in our network. Each node, edge and the whole graph are associated to a state vector. To build each state vector, we apply a multi-layer perceptron to the features of each considered object.

### S11.2.2 Graph update operation

The graph update operation takes a graph as input and outputs the same graph structure with updated state vector values. It consists of three steps :

1. update the edge state vectors

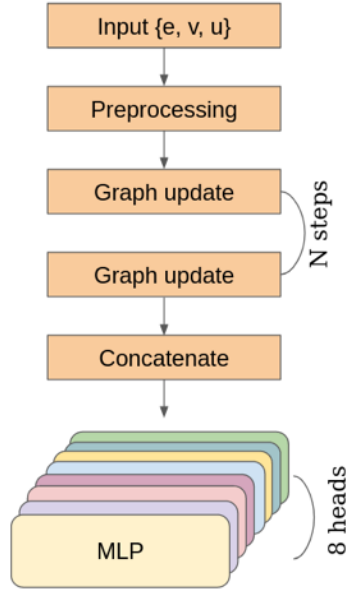

Figure S13: Best single model architecture

2. update the node state vectors

3. update the global state vector

We denote $V = \{\mathbf{v}_i\}_{i=1:N_v}$ the set of node vectors, with $N_V$ the node count, $E = \{\mathbf{e}_k\}_{k=1:N_E}$ the set of edge vectors, with $N_E$ the edge count, and $\mathbf{u}$ the global graph vector. $G = (V, E, \mathbf{u})$ is the input graph representation and $G' = (V', E', \mathbf{u}')$ is the output graph representation of the graph update operation.

In the following, we denote all $\phi$ functions as multi-layer perceptron with two hidden layers, SoftPlus activation and LayerNormalization that takes a vector as input and outputs another vector. If two $\phi$ functions share the same subscript it means that the multi-layer perceptrons share their parameters. If the $\phi$ function has no subscript it means that its parameters are not shared with any other multi-layer perceptron.

All $\psi$ functions are vector aggregation functions that outputs one vector from a set of vectors. In our architecture, all $\psi$ aggregations are attention functions that takes two parameters :

- a set of vectors to aggregate, which is used as the keys and values in the attention mechanism

- a vector which is used as query in the attention mechanism

The $\bigoplus$ operator is the vector concatenation operator.

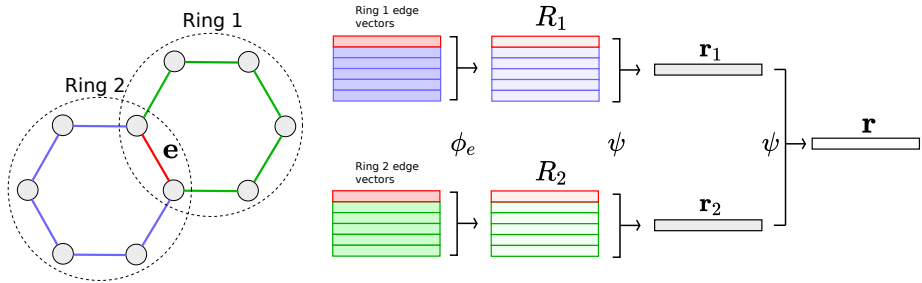

Figure S14: Computation of **r** with two rings

**Edge vector update** The edge vector update consists of four operations. In the following we denote **e** the edge vector to be updated and $b$ the bond associated with it.

Firstly we introduce an intermediary vector **r** which is an aggregation of the edge vectors contained in all rings $b$ is a part of, where a ring denote a simple cycle of atoms and chemical bonds in a molecule. The purpose of **r** is to allow a better flow of the rings related information. We proposed this intermediate vector because we observed that rings features had an high impact on preliminary scalar coupling models.

We denote $N_R$ the number of rings containing $b$ in the molecule. For the $i$th ring containing $b$, $R_i = \{\phi_e(\mathbf{e}_k)\}_{k=1:N_{R_i}}$ is the processed edge vectors of the ring. The vector **r** is computed as follows :

$$\mathbf{r}_i = \psi\Big(R_i, \phi_e(\mathbf{e})\Big)$$
$$\mathbf{r} = \psi\Big(\{\mathbf{r}_i\}_{i=1:N_R}, \phi_e(\mathbf{e})\Big)$$

$\mathbf{r}_i$ is an aggregated edge vector along the $i$th ring containing $b$ and **r** is an aggregated edge vector along in all the rings containing $b$. An example of such computation is displayed in Figure S14.

Secondly we introduce another intermediary vector **a** which is an aggregation of neighboring edge vectors and local geometric features. The purpose of **a** is firstly to allow to integrate angular edge-edge features into our architecture and secondly to provide a better flow of the edge-edge information. We proposed this intermediate vector because we observed that the engineered edge feature "angles between one edge and the edges formed with the closest atom" feature had an high impact on our preliminary scalar coupling models.

We denote $A = \{(\mathbf{e}_i^1, \mathbf{e}_i^2)\}_{i=1:N_V-2}$ is all couples of edges formed between an atom in the molecule and the two atoms in **e**. $\mathbf{f}_i$ is the feature vector characterising the triangle of edges $(\mathbf{e}, \mathbf{e}_i^1, \mathbf{e}_i^2)$ which is 5-dimensional and contains the 3 angles in the triangle as well as the length of edges $\mathbf{e}_i^1$ and $\mathbf{e}_i^2$. The vector **a** is defined as follows :

$$\mathbf{a} = \psi\Big(\Big\{\phi(\mathbf{f}_i)\bigoplus\phi_e(\mathbf{e})\bigoplus\phi_e(\mathbf{e}_i^1)\bigoplus\phi_e(\mathbf{e}_i^2)\Big\}_{i=1:N_V-2}, \phi_e(\mathbf{e})\Big)$$

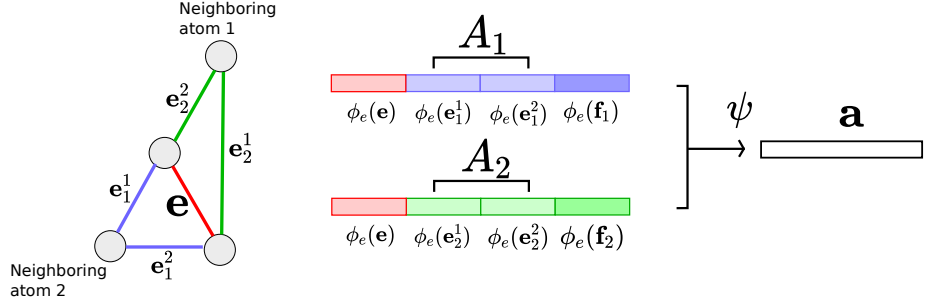

Figure S15: Computation of $\mathbf{a}$ for the edge $\mathbf{e}$ in a four node system.

An example of such computation is displayed in Figure S15.

Finally we compute the updated edge vector $\mathbf{e}'$. We denote the $\mathbf{v}_1$ and $\mathbf{v}_2$ the first and second atom vector of $\mathbf{e}$.

$$\mathbf{e}' = \phi\Big(\mathbf{r}\bigoplus \mathbf{a} \bigoplus \phi_e(\mathbf{e}) \bigoplus \phi_u(\mathbf{u})\Big) + \phi_{atom}(\mathbf{v}_1) + \phi_{atom}(\mathbf{v}_2) + \mathbf{e} \qquad \text{(S16)}$$

$\mathbf{e}'$ contains a skip connection to enable deeper and faster model training.

**Node update** We denote $\mathbf{v}$ the node vector to be updated, $\mathbf{v}_i$ the $i$th node vector different from $\mathbf{v}$ and $\mathbf{e}'_i$ the updated edge vector between the nodes associated with $\mathbf{v}$ and $\mathbf{v}_i$. Let $C = \{\mathbf{v}_i \bigoplus \mathbf{e}'_i\}_{i=1:N_V-1}$ be the set of concatenated node and updated edge vectors. The vector $\mathbf{v}'$ is defined as follows :

$$\mathbf{v}_{agg} = \psi\Big(C, \phi_v(\mathbf{v})\Big)$$
$$\mathbf{v}' = \phi_s\Big(\mathbf{v}_{agg} \bigoplus \phi_v(\mathbf{v}) \bigoplus \phi_u(\mathbf{u})\Big) + \mathbf{v}$$

As for the edge update, we integrate a skip connection.

**Global state update** We denote $V' = \{\mathbf{v}'_k\}_{k=1:N_V}$ the set of updated node vectors and $E' = \{\mathbf{e}'_k\}_{k=1:N_E}$ the set of updated edges vectors. The updated global state vector is defined as follows :

$$\mathbf{u}_{agg\_edge} = \psi\Big(E', \phi_u(\mathbf{u})\Big)$$
$$\mathbf{u}_{agg\_node} = \psi\Big(V', \phi_u(\mathbf{u})\Big)$$
$$\mathbf{u}' = \phi\Big(\mathbf{u}_{agg\_edge} \bigoplus \mathbf{u}_{agg\_node} \bigoplus \phi_u(\mathbf{u})\Big) + \mathbf{u}$$

As for the edge and node update, we integrate a skip connection.

### S11.2.3 Readout stage

To predict a scalar coupling associated with an edge, we concatenated the edge vector, the two node vectors associated with the edge, and the global state vector and passed it through multi-layer perceptron to generate the final output. Since there was 8 different types of scalar coupling, our model had 8 different multi-layer perceptrons to enforce a different readout for each type.

## S11.3 Training and ensembling

We trained our model with the Adam[20] optimizer with a fixed learning rate and the original article default parameters, for about 150 epochs then reduced the learning rate by a factor 2 each 3 epochs for 15 epochs.

The dimension of our graph vectors was set at 300. Our batch size was 20, a relatively small number due to the high memory requirements of the computation of the neighboring edge vectors aggregation.

### S11.3.1 Experiments and model variations

We observed that even models with a modest score could help to contribute to a better overall performance with an ensemble model. As such we tried various architectures with different modifications. The following experiments were tested and integrated as base models:

- Different activation functions in our multi-layer perceptron: Softplus provided a boost of performance in comparison to a ReLU baseline.

- Normalization: LayerNormalization worked better that BatchNormalization in our experiments and helped improve convergence.

- In the readout stage, rather that using only the edge and two nodes associated with a scalar coupling, we concatenated all the edges and nodes vectors in the chemical bond path of the two atoms we compute the scalar coupling. We observed faster training with this method but could not integrate it in our best single model in time.

- We iterated with different number of graph update operations or different number of hidden layers.

### S11.3.2 Ensemble learning

Ensemble learning is a technique that aggregates multiple models into a single one to obtain a better performance. It has become a standard in machine learning competitions. To build our final prediction, we fitted a linear model and a gradient boosting model that we averaged. Rather than using only our 16 base models in this ensemble, we also integrating intermediary models checkpoints of those 16 models to further improve the performance. In the end there was about 50 input models in the ensemble.

### S11.4  Improvements

We can think of few improvements for our best single model architecture:

- Prune the amount of edges considered in the neighboring edge vectors aggregation by keeping only the edges closest to the updated edge. This would greatly reduce the memory consumption of the architecture, allow a bigger batch size and fasten the model. We suppose that this would not degrade the performance as the angular features that had high permutation importance in our preliminary models were related to the closest edges.

- Integrate the full chemical bond path in the readout stage mentioned in S11.3.1 to increase the training speed as it proved efficient on other similar architectures.

## S12  Solution 12

### S12.1  Introduction

There has been extensive work recently using graph neural networks for predicting properties of molecular systems. Many problems along this direction require the use of configuration information (i.e., the positions of all atoms in a molecule.) A key challenge in applying machine learning techniques to these problems is that of capturing local geometric information (such as bond or torsion angles), while preserving symmetries of the overall system. Symmetries such as rotation and translation invariance are fundamental properties of molecular systems, and therefore must be exactly preserved in order for a machine learning architecture to effectively use training data and make meaningful predictions.

Our recent paper proposed the COvaRiant MOleculaR Artificial Neural neTwork *(Cormorant)* to help solve these issues [42]. *Cormorant* uses spherical tensors as activation to encode local geometric information around each atom's environment. A spherical tensor is an object $F^\ell$ which transforms in a predictable way under a rotation. Specifically, for a rotation $R \in \text{SO}(3)$, there is a matrix $D^\ell(R)$, such that $F^\ell \to D^\ell(R) F^\ell$. Here, $D^\ell(R)$ is known as a Wigner-D matrix, and $\ell$ is known as the order of the representation.

A tensor with $\ell = 0$ is a scalar quantity (that is, invariant under rotations), whereas $\ell = 1$ is a vector quantity such as a dipole, and $\ell = 2$ is a second-order tensor like a quadrupole, and so on. This structure is well known in physics. For example, the multipole expansion is a decomposition of an electrostatic potential $V(\mathbf{r}) = \sum_{\ell=0}^{\infty} Q^\ell Y^\ell(\hat{\mathbf{r}})/r^{\ell+1}$ into multipole moments $Q^\ell$ and spherical harmonics $Y^\ell(\hat{\mathbf{r}})$ oriented in the direction of the vector $\mathbf{r}$. For a more in-depth discussion of spherical tensors, we point the reader to Ref. [42].

The use of spherical tensors allows for a network architecture that is "covariant to rotations." This means that if level $s$, a rotation is applied to all activations $F^{s,\ell} \to D^\ell(R) F^{s,\ell}$, then at the next level $s+1$, all activations will automatically inherit that rotation so $F^{s+1,\ell} \to D^\ell(R) F^{s+1,\ell}$. As such, a rotation to the input of *Cormorant* will propagate through the network to ensure the output transforms as well. In this way, we can capture local geometric information, but can still maintain the desired transformation properties under rotations.

A key point is that the quantum algebra (SU(2)) used in the definition of NMR couplings is for our purposes equivalent (homomorphic) to the SO(3) algebra used in *Cormorant*[1]. The $J$-Coupling Hamiltonian

$$H = 2\pi \mathbf{I}_i \cdot \mathbf{J}_{ij} \cdot \mathbf{I}_j$$

describes the couplings $\mathbf{J}_{ij}$ between spin operators $\mathbf{I}_i$. The spin operators are themselves objects that generate the Lie algebra $SU(2)$, and thus transform according to Wigner-D matrices. *Cormorant* therefore naturally incorporates the structure of $J$-Coupling Hamiltonian, and we expect is a natural platform learning NMR couplings.

### S12.2 Model architecture

We now briefly summarize our *Cormorant* architecture; for a more in-depth description of our architecture, see the original paper [42].

The structure of of *Cormorant* is similar to a graph neural network, with the key difference that vertex activations $F_i$ and edge activations $G_{ij}$ for atoms $i, j$ are promoted to lists of spherical tensors $F_i = \left(F_i^0, \ldots, F_i^{\ell_{\max}}\right)$ and $G_{ij} = \left(G_{ij}^0, \ldots, G_{ij}^{\ell_{\max}}\right)$, where $F_i^\ell \in \mathbb{C}^{(2\ell+1) \times n_\ell}$ and $G_{ij}^\ell \in \mathbb{C}^{(2\ell+1) \times n_\ell}$ are spherical tensors of order $\ell$, and $n_\ell$ is the multiplicity (number of channels) of the tensor.

In order to maintain covariance, we must carefully choose our non-linearity. The core non-linearity in *Cormorant* is dictated by the structure of the $D^\ell(R)$ matrices, and is known as the Clebsch-Gordan product [CG]. We express the CG product of $\otimes_{\text{cg}}$ of two spherical tensors as:

$$[A_{\ell_1} \otimes_{\text{cg}} B_{\ell_2}]_\ell = \bigoplus_{\ell=|\ell_1-\ell_2|}^{\ell_1+\ell_2} C_{\ell_1 \ell_2 \ell} (A_{\ell_1} \otimes B_{\ell_2})$$

where $\otimes$ denotes a Kronecker product, and $C_{\ell_1 \ell_2 \ell}$ are the famous Clebsch-Gordan coefficients [CG].

Our vertex activations $F_i^s$ at level $s$ are chosen to be

$$F_i^s = \left[ F_i^{s-1} \oplus \left( F_i^{s-1} \otimes_{\text{cg}} F_i^{s-1} \right) \oplus \left( \sum_j G_{ij}^s \otimes_{\text{cg}} F_j^{s-1} \right) \right] \cdot W_{s,\ell}^{\text{vertex}}$$

where $\oplus$ denotes concatenation and $W_{s,\ell}^{\text{vertex}}$ is a linear mixing layer that acts on the multiplicity index. We choose the edge activations to have the form of $G_{ij}^{s,\ell} = g_{ij}^{s,\ell} \times Y^\ell(\hat{\mathbf{r}}_{ij})$, where $\mathbf{r}_{ij}$ is a vector pointing from atom $i$ to atom $j$. The scalar-valued edge terms are given by

$$g_{ij}^{s,\ell} = \mu^s(r_{ij}) \left[ \left( g_{ij}^{s-1,\ell} \oplus \left( F_i^{s-1} \cdot F_j^{s-1} \right) \oplus \eta^{s,\ell}(r_{ij}) \right) \cdot W_{s,\ell}^{\text{edge}} \right]$$

with $\mu^s(r_{ij})$ a learnable mask function, $\eta^{s,\ell}(r_{ij})$ a learnable set of radial basis functions, and $W_{s,\ell}^{\text{edge}}$ a linear layer along the multiplicity index.

---

[1] More precisely, SU(2) is a double cover of SO(3), and both groups share a Lie algebra.

We iterate this architecture for $s = 0, \ldots, s_{\max}$. Finally, at the last layer of *Cormorant*, we take the $\ell = 0$ component of the edge $g_{ij}^{s_{\max}\ell}$, and use it as a prediction of the J-coupling.

### S12.3 Training and ensembling

To predict J-coupling constants, we first observed that each J-coupling was a rotationally invariant feature on pairs of atoms. Consequently, it was natural to read the values of the J-couplings off the values of $g_{ij}^{s,\ell}$. We therefore constructed a neural net with five cormorant layers, with a maximum $\ell$ value of 3. In each layer, we used 48 vertex activations for each value of $\ell$. We then took a linear combination of the values of $g_{ij}^{s,\ell}$ for each layer and took a learned linear combination of the values as our prediction for the scalar coupling constant. Initial features were constructed using the radial distance between atoms, the atomic identity, and the atomic charge. In general, almost exactly the same network was used as for the QM9 tests in the Cormorant paper [42]. We also used the Mulliken charges as initial features for pretraining. Our training proceeded in three stages.

1. Three nets were trained on the 1J, 2J, and 3J couplings with the Mulliken Charges on an internal split on the training data.

2. Eight nets were trained on the 1JHC, 1JHN, 2JHH, 2JHC, etc. couplings with the Mulliken Charges on an internal split on the training data, with weights initialized to be the values in stage 1.

3. Each of the nets in stage 2 was trained on the full dataset, without access to the Mulliken charges.

The internal split was constructed using an 80/10/10 train/validation/test split on the training dataset provided in the Kaggle competition. The model was optimized using the AMSGrad optimizer for 200 epochs. The learning rate was varied using cosine annealing. For stage one the initial and final learning was $5 * 10^{-4}$ and $5 * 10^{-6}$, and for stages two and three the initial and final learning rates were $3 * 10^{-4}$ and $3 * 10^{-6}$, respectively. We repeated stages two and three with different random seeds in attempt to average over multiple trained networks, however this had negligible effect on the results.

In table S6, we give the final results on the Kaggle Test/Train split. We achieve considerably stronger results for the 2J and 3J couplings than for the 1J couplings. However, we were pleasantly surprised by the minimal fine-tuning required to get results close to state of the art. Future directions include directly connecting the tensors learned by cormorant with the tensor-valued data in NMR experiments to provide more detailed inferences.

Table S6: Performance

| Dataset | 1JHC | 1JHN | 2JHH | 2JHC | 2JHN | 3JHH | 3JHC | 3JHN |
|---|---|---|---|---|---|---|---|---|
| Training | -2.736 | -4.834 | -4.366 | -3.481 | -5.4889 | -4.632 | -3.275 | -5.557 |
| Test | -1.768 | -2.224 | -3.584 | -2.856 | -3.312 | -3.552 | -2.688 | -3.496 |

## S13 Code and data

Code and data is available on http://osf.io/kcaht and http://github.com/larsbratholm/champs_kaggle. These contain the following

- List of molecules removed as they did not contain hydrogens.
- List of molecules removed due to one or more couplings being outliers.
- List of training and test molecules.
- Script to convert QM9 extended XYZ-files into both regularly formatted XYZ-files as well as input files for Gaussian NMR computations.
- Script to create the Kaggle dataset from the output files of Gaussian NMR computations.
- Archive of all extended XYZ-files of QM9 that passed the consistency check of the original paper.
- Archive of all regular formatted XYZ files parsed from the extended XYZ file format of the QM9 dataset.
- Archive of all Gaussian input files.
- Archive of all Gaussian output files.
- Archive of the Kaggle dataset.
- Archive of the top 400 submissions.
- Script to create an ensemble from the top submissions.
- Script to create the plots used in the paper and SI.
- Code used by the 1st, 2nd, 3rd, 4th, 5th and 12th placed teams to create their models.